\documentclass[10pt,prx,aps,showpacs,twocolumn,preprintnumbers,amsmath,amssymb,superscriptaddress,longbibliography]{revtex4-2}
\usepackage{graphicx}    
\usepackage{latexsym}    
\usepackage{color}       
\usepackage{dcolumn} 
\usepackage{bm}          
\usepackage{amssymb}   
\expandafter\let\csname equation*\endcsname\relax
\expandafter\let\csname endequation*\endcsname\relax
\usepackage{amsmath}
\usepackage{mathtools}  
\usepackage{bbm}
\usepackage{graphics}
\usepackage{epsfig}
\usepackage{subfigure}
\usepackage{verbatim}
\usepackage{enumerate}
\usepackage{footnote}
\usepackage[svgnames]{xcolor}
\renewcommand{\arraystretch}{1.8}
\usepackage{xr-hyper}
\usepackage[colorlinks=true,allcolors=MediumBlue]{hyperref}  
\usepackage{orcidlink}

\begin{document}

\begin{abstract}
Sampling synthetic turbulent fields as a computationally tractable surrogate for direct numerical simulations (DNS) is an important practical problem in various applications, and allows to test our physical understanding of the main features of real turbulent flows. Reproducing higher-order Eulerian correlation functions, as well as Lagrangian particle statistics, requires an accurate representation of coherent structures of the flow in the synthetic turbulent fields. To this end, we propose in this paper a systematic coherent-structure based method for sampling synthetic random fields, based on a superposition of instanton configurations -- an instanton gas -- from the field-theoretic formulation of turbulence. We discuss sampling strategies for ensembles of instantons, both with and without interactions and including Gaussian fluctuations around them. The resulting Eulerian and Lagrangian statistics are evaluated numerically and compared against DNS results, as well as Gaussian and log-normal cascade models that lack coherent structures. The instanton gas approach is illustrated via the example of one-dimensional Burgers turbulence throughout this paper, and we show that already a canonical ensemble of non-interacting instantons without fluctuations reproduces DNS statistics very well. Finally, we outline extensions of the method to higher dimensions, in particular to magnetohydrodynamic turbulence for future applications to cosmic ray propagation.
\end{abstract}

\title{Synthetic Turbulence via an Instanton Gas Approximation}

\author{Timo Schorlepp\,\orcidlink{0000-0002-9143-8854}}
\email{timo.schorlepp@nyu.edu}
\affiliation{Courant Institute of Mathematical Sciences, New York University,
New York, NY, USA}
\author{Katharina Kormann\,\orcidlink{0000-0003-1956-2073}}
\email{k.kormann@rub.de}
\affiliation{Numerical Mathematics, Ruhr University Bochum,
Bochum, Germany}
\author{Jeremiah L\"ubke\,\orcidlink{0000-0001-6338-9728}}
\email{jeremiah.luebke@rub.de}
\affiliation{Institute for Theoretical Physics I, Ruhr University Bochum,
Bochum, Germany}
\author{Tobias Sch\"afer\,\orcidlink{0000-0002-9255-4718}}
\email{tobias.schaefer@csi.cuny.edu}
\affiliation{Department of Mathematics, College of Staten Island,   
Staten Island, NY, USA \& Physics Program, CUNY Graduate Center, 
New York, NY, USA}
\author{Rainer Grauer\,\orcidlink{0000-0003-0622-071X}}
\email{grauer@tp1.rub.de}
\affiliation{Institute for Theoretical Physics I, Ruhr University Bochum,
Bochum, Germany}

\date{\today}

\maketitle

\section{Introduction}
\label{sec:intro}
Synthetic turbulence generation -- sampling random fields which mimic a turbulent state, but without actually solving the underlying equations of motion of the fluid -- has been an active field of research for 50 years~\cite{kraichnan:1970}, driven by its wide-ranging applications across various disciplines.
One motivation for this line of research is the need to construct surrogates for turbulent fields, required as an ingredient for a variety of applications, that are computationally cheaper to generate than full direct numerical simulations (DNS), but still display pertinent characteristics of turbulence. A prominent example is the synthesis of wind fields with accurate turbulent fluctuations, which plays a crucial role in optimizing offshore wind farm performance and determining the maximum allowable stresses on turbine blades~\cite{kleinhans-etal:2010,chabaud:2024}. In the same vein, synthetic turbulence is also used to generate inflow boundary conditions and model Reynolds stresses in large eddy simulations~\cite{smirnov-shi-celik:2001,jarrin-benhamadouche-laurence-etal:2006,montomoli-eastwood:2011,bode-gauding-etal:2021}.

In addition to its practical importance in these applications, generating synthetic turbulence plays a significant role in deepening our physical understanding of the fundamental mechanisms of real turbulence. By building reduced models and mimicking key features of turbulent flows, synthetic turbulence helps to isolate and understand the role of individual coherent structures and their interplay~\cite{novikov:1983,cheklov-yakhot:1995,hatakeyama-kambe:1997,kambe-hatakeyama:2000,wilczek-jenko-friedrich:2008,apolinario-moriconi-pereira-etal:2020,durrive_SwiftGeneratorThreedimensional_2022,zinchenko-pushenk-schumacher:2024,shen-yao-yang:2024,moriconi-pereira-valadao:2024,maci-keppens-bacchini:2024,ruffenach-fery-dubrulle:2025,han-shen-yang:2025}, a central component of the ``turbulence problem''~\cite{sreenivasan-schumacher:2025}. 

In this work, by adopting ideas from quantum and statistical field theory, we propose and explore a new and physically principled approach for synthetic turbulence generation.

\begin{figure*}[ht!]
    \centering
    \includegraphics[width=\linewidth]{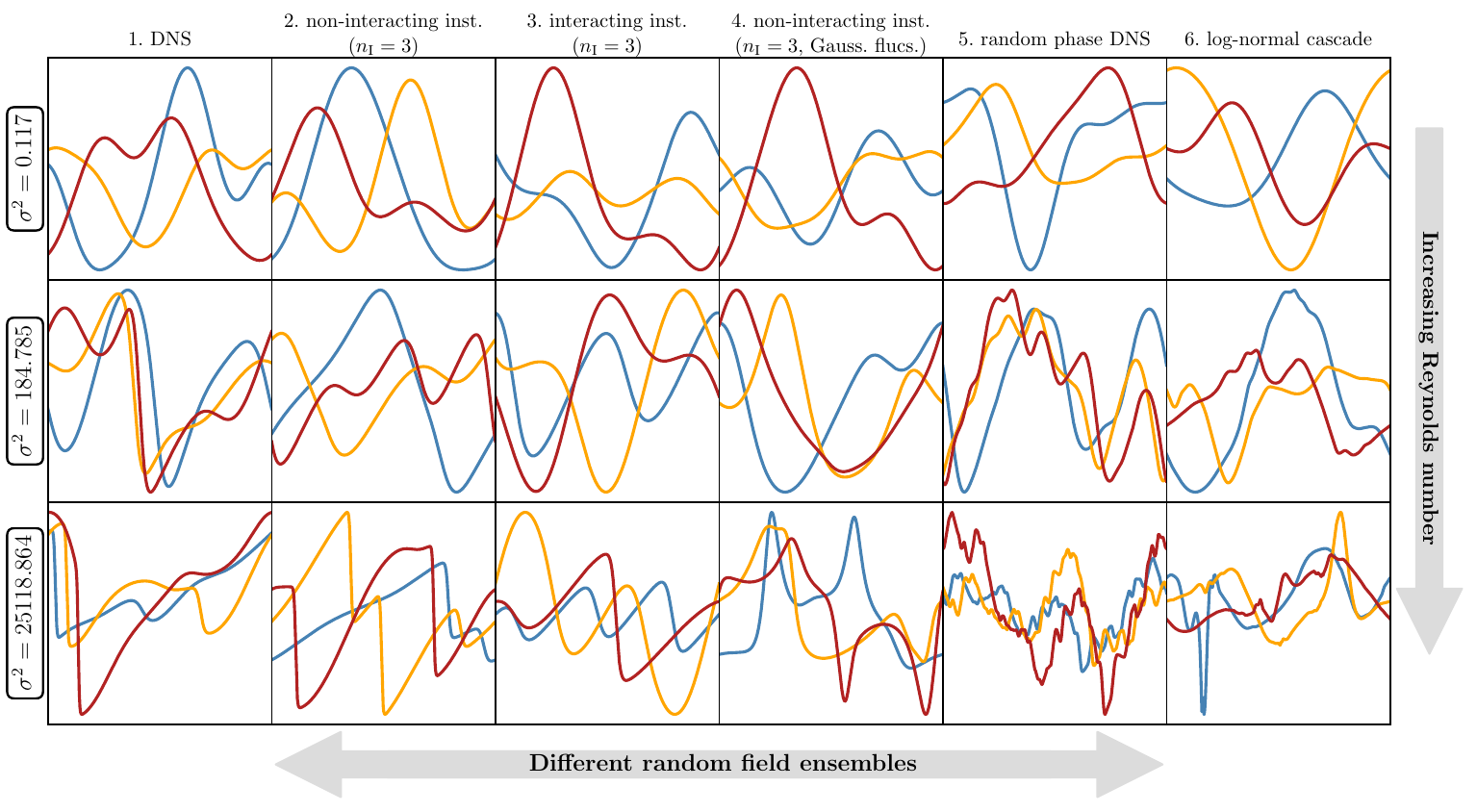}
    \caption{Three independent snapshots of random velocity field realizations $u(x)$, $x \in [0, 2 \pi]$, for each of the six random field ensembles analyzed in Sec.~\ref{sec:results} (columns), for three different noise variances $\sigma^2 \in \{1.17 \cdot 10^{-1}, 1.85 \cdot 10^2, 2.51 \cdot 10^4 \}$ (rows), corresponding to Reynolds numbers $\text{Re} \in \{1.03, 32.29, 192.21\}$ for DNS of the Burgers Eq.~\eqref{eq:burgers}. Technical details are explained in Sec.~\ref{sec:inst-intro} to~\ref{sec:other-ensembles}, and statistical results are analyzed in Sec.~\ref{sec:results}. The Burgers DNS fields (column~1) display stronger and stronger negative gradients or quasi-shocks, as well as smooth ramps, as the Reynolds number increases. This is reproduced to varying degrees by the instanton gas ensembles (columns~2 to~4), but not by the structureless synthetic fields (columns~5 and~6).}
    \label{fig:snapshots}
\end{figure*}

A practical motivation for this is the modeling of synthetic magnetohydrodynamic (MHD) turbulence in astrophysical environments to understand the propagation of cosmic rays in turbulent magnetic fields~\cite{strong:2007,tjus-merten:2020,dundovic-etal:2020,lazarian:2023}, to learn about the impact of turbulence, intermittency and singular structure on e.g.\ diffusion constants. Another key aspect is identifying the conditions under which particle transport behaves diffusively, sub- or super-diffusively.
The need to generate synthetic turbulence in astrophysical systems arises from the vast distances cosmic rays travel and the extremely high numerical resolutions required, which will remain impractical to treat via DNS in the foreseeable future.

Specifically, in this setting, spontaneous alignment between the fluid flow and magnetic field in (nearly) collisionless plasmas leads to the formation of coherent, nearly singular structures (i.e.\ current sheets), which are long-lived, exhibit intense amplitudes and break the scale-invariance of the system. This leads to strongly non-Gaussian distributions for fluctuations and anomalous scaling of structure functions, which is an indication of intermittency~\cite{biskamp:2003}. There is a long-standing history of generating turbulent fields with specified intermittency properties, during which a wide range of methods were used (see e.g.\ Refs.~\cite{
meneveau-sreenivasan:1987,
juneja-lathrop-etal:1994,
biferale-bofetta-etal:1998,
rosales-meneveau:2008,
alouanibibi-leroux:2014,
subedi-chhiber-etal:2014,
pucci-malara-etal:2016,
wilczek:2016,
pereira-garban-chevillard:2016,
muzy:2019,
reneuve-chevillard:2020,
sinhuber-etal:2021,
luebke-friedrich-grauer:2023,
luebke-effenberger-etal:2024,
li-biferale-etal:2024,
martin2025,
pisoni-peddinti-tiunov-etal:2025,
lesaffre-durrive-goossaert-etal:2025} and references therein).
However, recently, in the context of MHD turbulence, the challenges in constructing a method which exhibits the desired features like intermittency were demonstrated in Ref.~\cite{luebke-effenberger-etal:2024}, where it was shown that a multifractal field, which correctly reproduces higher-order scaling, does \textit{not} significantly affect charged particle transport and that the incorporation of simple advective structures into the model on the one hand leads to improved diffusion coefficients, but on the other hand changes the scaling behavior disproportionately away from the expectations.

These findings emphasize the statement that nearly singular structures such as shocks, vortices or sheets can cause intermittency, but the reverse is not true: intermittency, expressed in anomalous scaling of structure functions, does not necessarily imply coherent near-singular structures. The details of nearly singular structures are embedded within higher-order multipoint correlations.
This motivates the synthesis of turbulent fields through the direct and systematic superposition of appropriate coherent structures, as has been done, e.g., in Refs.~\cite{hatakeyama-kambe:1997, kambe-hatakeyama:2000,wilczek-jenko-friedrich:2008,zinchenko-pushenk-schumacher:2024} using Burgers vortices as building blocks for synthetic hydrodynamic turbulence. 
These studies gave remarkable results for statistical quantities, but with a method of superposing such structures that does not follow from first principles. Moreover, in the context of MHD turbulence, no analogous solutions exist to Burgers vortices in the Navier--Stokes equations, thereby underscoring the need for a systematic approach.

In the present work, we attempt to introduce a substantial change to the way we choose the coherent structures for the synthesis of turbulence, by not relying on a phenomenologically motivated choice, such as Burgers vortices for synthetic Navier--Stokes turbulence, but instead using those coherent structures that naturally appear in the field theoretic formulation of turbulence~\cite{gurarie-migdal:1996,falkovich:1996,balkovsky-falkovich-kolokolov-etal:1997,grafke-grauer-schaefer:2015c} and can be systematically computed from the governing equations. These field configurations, termed instantons, give the largest contribution to the functional integral that determines the statistics of observables of interest in the system. In a certain sense, they can be regarded as the prototypical singular structures that arise in turbulence, and hence building synthetic fields out of them is a natural idea. In addition, these solutions can be superimposed as an instanton gas/fluid by rules governed by the underlying path integral formulation.

This defines the aim of this work, which is to generate synthetic turbulence by using a superposition of instanton configurations, the result of which we will term an ``instanton gas'', and to investigate the statistical and physical properties of the resulting synthetic fields. In other words, we want to answer the question whether an instanton gas can be used as an accurate representation of a true turbulent field, both in terms of Eulerian field statistics (energy spectrum, spatial increment statistics with meaningful intermittency exponents, and corresponding non-Gaussian probability distributions for fluctuations), as well as regarding the emerging coherent structures that ensure correct Lagrangian statistics for particles transported by the synthetic turbulent fields.

This instanton-based approach to synthetic turbulence is desirable for two reasons: (i) it is based on well-defined approximations of the underlying path integral formulation of turbulence, and provides clear target distributions to sample from that are physically interpretable (ii) instanton configurations can be precomputed in an ``offline phase'', and online samples of the instanton gas for applications can be generated cheaply. We note that similar instanton-gas or liquid based reduced models constitute well-established approximations in quantum chromodynamics~\cite{schaefer-shuryak:1996,schaefer-shuryak:1998,schaefer:2004} and solid state physics~\cite{ulybyshev-winterowd-assaad-etal:2023}, but, to the best of our knowledge, have not been used to construct synthetic turbulent fields before.

As for potential limitations of the instanton-gas method, we first point out that recent advances in machine learning provide an alternative, and practically successful, route to turbulence generation, e.g.\ via generative adversarial networks~\cite{drygala-winhart-dimare-etal:2022, drygala-dimare-gottschalk:2023} or diffusion models~\cite{lienen-luedke-hansen-palmus-etal:2023,saydemir-lienen-guennemann:2024,du-parikh-fan-etal:2024,gao-han-fan-etal:2024}. Nevertheless, we aim for a more physically principled and interpretable synthetic turbulence simulation here, which retains its merits despite those advances. Furthermore, computing the necessary instanton configurations for our method numerically through optimization can remain a technical challenge to implement. However, recently, progress has been made to develop efficient and manageable approaches, see e.g.\ Refs.~\cite{schorlepp-grafke-may-etal:2022,zakine-vanden-eijnden:2023,simonnet:2023,schorlepp-grafke:2025}. Lastly, as we discuss in more detail below, there are a few to a certain extent arbitrary -- but physically motivated -- parameters in the method implemented in this paper, such as the choice of observable, and the number of instantons for the canonical ensemble, that may require some fine-tuning. Still, compared to sampling structured random fields using other parametric families of fields with similar methods as discussed in Sec.~\ref{sec:inst-gas}, the instanton-based approach could be regarded as more canonical, as it removes a lot of the arbitrariness and reduces the number of free parameters.
However, we would like to point out that an instanton gas approximation of higher-dimensional and complex systems still raises many open questions. The question of suitable observables must be physically motivated. In the case of the Burgers equation discussed next, this is natural, but in the case of the three-dimensional Navier--Stokes equations, for example, the question must then determine whether vorticity or strain is chosen. In the example of MHD turbulence shown in the outlook, the current density is chosen as the observable, since an ensemble of current sheets plays a central role in the diffusion of cosmic rays.

In this paper, for simplicity and concreteness, we illustrate the approach using one-dimensional Burgers turbulence as an example, and we will demonstrate excellent agreement of the aforementioned statistical properties of the synthesized instanton gas fields in this system with those obtained from DNS of the Burgers equation as the ground truth. For a more complete picture, we will further compare the statistical properties of the instanton gas ensembles of this paper to two other random field ensembles without coherent structures -- which are shocks in the case of the Burgers equation -- to isolate the influence of those structures. On the one hand, we will consider phase-randomized (in Fourier space) versions of the DNS fields~\cite{koga-chian-miranda-etal:2007,grauer-homann-pinton:2012,shukurov-snodin-seta-etal:2017}, and on the other hand a synthetic turbulence model with a log-normal cascade from Ref.~\cite{muzy:2019}. We show example snapshots of samples from the different random field ensembles considered in this paper -- from DNS of the Burgers equation, different instanton gas ensembles, and the two ensembles without coherent structures -- in Fig.~\ref{fig:snapshots}. Details will be explained in subsequent sections. We also note here that a related coherent structure-based approach  for modeling Burgers turbulence can be found in Ref.~\cite{cheklov-yakhot:1995}, but without relying on instanton solutions.

The Burgers equation~\cite{burgers:1948}, introduced originally as a toy model for Navier--Stokes turbulence, is a commonly studied and strongly intermittent turbulent system with various applications in its own right~\cite{frisch-bec:2002,bec-khanin:2007}. It reads
\begin{equation}
\partial_{t} u + u \partial_x u -\nu \partial_{x x}u = \chi^{1/2} * \eta \; ,
\label{eq:burgers}
\end{equation}
where $*$ denotes convolution in space. By rescaling the velocity field $u$ and time $t$ we set $\nu=1$ without loss of generality and control the turbulence level (and thus the Reynolds number $\text{Re}$) through the magnitude of the forcing. The Burgers Eq.~\eqref{eq:burgers} is considered in a periodic box of size $L_{\text{box}} = 2 \pi$. The Gaussian random forcing~$\eta$ with $\mathbb{E}\left[\eta(x,t) \right] = 0$ is $\delta$-correlated in space and time $\mathbb{E}\left[\eta(x,t) \eta(x^\prime,t^\prime)\right] = \sigma^2 \delta(x-x^\prime) \delta(t-t^\prime)$ with variance $\sigma^2 > 0$. The correlation function $\chi = \chi^{1/2} * \chi^{1/2}$ is invariant under translations $\chi(x,x')=\chi(x-x')$. We consider the typical choice of a ``Mexican hat'' correlation function $\chi(x) = - \partial_{xx} \left( e^{-x^2/2} \right)$ for concreteness, which corresponds to an example of a spatially smooth forcing of the turbulence on large scales. In line with the classical cascade picture of turbulence~\cite{bec-khanin:2007,richardson:2007,frisch:1995}, the nonlinearity of Eq.~\eqref{eq:burgers} then leads to the formation of shocks with large negative gradients, which are regularized and dissipated through viscosity.

The outline of the paper is as follows. First, we briefly recall the necessary elements of the instanton formalism to calculate turbulence statistics in Sec.~\ref{sec:inst-intro}.
Numerical details and results of DNS of the Burgers Eq.~\eqref{eq:burgers}, as well as the single-instanton formalism in itself, are presented in Sec.~\ref{sec:num-details}.
Subsequently, we derive the instanton gas approximation used for synthesizing turbulent fields in Sec.~\ref{sec:inst-gas}. We will then discuss our Monte Carlo approach to sample from different instanton gas ensembles in Sec.~\ref{sec:mc}, where the main ingredients are the one-point probability density function (PDF) $\rho(a) = \mathbb{E} \left[\delta(\partial_x u(x_0, t_0) - a) \right]$ of the gradients $\partial_x u$ in the turbulent field, a Gram determinant~$g$, and possible interactions~$\Delta S$ between instantons which enter the steps of a Markov chain Monte Carlo (MCMC) algorithm. Other random field ensembles without coherent structures for comparison are briefly introduced in Sec.~\ref{sec:other-ensembles}. We will then present numerical results for the different random field ensembles on Eulerian statistics — PDFs of gradients, energy spectra, and structure functions — as well as particle statistics, which exhibit similarities to the characteristics of cosmic ray propagation in turbulent magnetic fields, in Sec.~\ref{sec:results}. Among the different possible instanton gas ensembles that are introduced theoretically in Sec.~\ref{sec:inst-gas}, the ones that we sample numerically for the results of this paper in Sec.~\ref{sec:inst-gas} are: a canonical ensemble of $n_{\text{I}} = 3$ non-interacting instantons without fluctuations around them (ensemble 2), a canonical ensemble of $n_{\text{I}} = 3$ interacting instantons, without approximations to their interaction $\Delta S$, and without fluctuations around them (ensemble 3), and a canonical ensemble of $n_{\text{I}} = 3$ non-interacting instantons with Gaussian fluctuations around the individual instantons (ensemble 4). We do not study simplified interaction models for $\Delta S$ or grand-canonical ensembles of instantons numerically in this paper, leaving this as future work. We conclude with a summary and discussion in Sec.~\ref{sec:concl}, and provide an outlook on the necessary steps to extend this method to MHD turbulence, ultimately enabling its application to cosmic ray propagation, in Sec.~\ref{sec:outl}. Appendix~\ref{app:muzy} contains additional details about the selection of hyperparameters for the log-normal cascade model.

\section{The instanton formalism}
\label{sec:inst-intro}
The instanton formalism~\cite{thooft:1976,coleman:1979,vainshtein-zakharov-novikov-etal:1982} was introduced to turbulence theory in the 1990s by Refs.~\cite{gurarie-migdal:1996,falkovich:1996,balkovsky-falkovich-kolokolov-etal:1997}, see also Ref.~\cite{grafke-grauer-schaefer:2015c} for a historical overview and context. It involves formulating turbulence statistics, e.g.\ the PDF of an observable such as the gradient $\partial_x u$ at one point in space and time $(x_0, t_0)$,  as a path integral and applying the saddle-point or Laplace method to approximate the path integral, either in the limit of small noise $\sigma \to 0$ or in the regime of large observable values $\lvert a \rvert \to \infty$. We note that the instanton approach is also known as large deviation theory~\cite{varadhan-1984,touchette:2009,freidlin-2012,grafke-vanden-eijnden:2019}.

The important steps of exactly including Gaussian fluctuations around the instanton, as well as possible zero modes, have been made practically feasible through a series of recent works~\cite{schorlepp-grafke-grauer:2021,bouchet-reygner:2022,schorlepp-grafke-grauer:2023,schorlepp-tong-etal:2023,grafke-schaefer-vanden-eijnden:2023,heller-limmer:2024,schorlepp-grafke:2025} (see also Ref.~\cite{apolinario-moriconi-pereira:2019} for an alternative perturbative approach for the Burgers Eq.~\eqref{eq:burgers}). In this paper, we follow the method introduced in Refs.~\cite{schorlepp-tong-etal:2023,schorlepp-grafke:2025}, which is especially well suited for problems with large scale forcing (cf.\ Ref.~\cite{burekovic-schaefer-grauer:2024}).

Our goal later will be to compose synthetic turbulence as a gas of instanton configurations $u^{\text{I}}_{a,x_0}(\cdot, t_0)$ at different gradient strengths~$a$ and positions~$x_0$. Consequently, we first need to be able to compute instantons at given gradient strengths~$a$, as well as the one-point gradient PDF $\rho(a)$. Note that because of translational invariance of Eq.~\eqref{eq:burgers}, we simply have $u^{\text{I}}_{a,x_0}(x, t_0) = u^{\text{I}}_{a,0}(x - x_0, t_0)$ here. More generally, beyond the example of Burgers turbulence, the approach presented in this paper hence encompasses choosing a representative observable of the system at hand. For MHD turbulence, this could e.g.\ be the current density, but other choices would be possible as well.

Now, to compute the instanton configurations and approximate the one-point gradient PDF, we adopt a path integral formulation of the stochastic process described by Eq.~\eqref{eq:burgers}, such as the one originally introduced by Onsager and Machlup~\cite{onsager-machlup:1953, machlup-onsager:1953} in terms of the velocity field~$u$, or its "linearized" version with an additional response field~$p$~\cite{janssen:1976,dedominicis:1976}. In the following, we will directly utilize the underlying formulation as a path integral over the Gaussian space-time white noise field $\eta$ instead~\cite{schorlepp-tong-etal:2023,schorlepp-grafke:2025}:
\begin{equation}
    \rho(a)=\int \mathcal{D} \eta \; \delta\left(F\left[\eta\right]-a \right) \exp \left\{- \frac{1}{2 \sigma^2} \|\eta\|_{L^2}^{2}\right\} \; .
    \label{eq:path_integral}
\end{equation}
Here, $\| \cdot \|_{L^2}$ denotes the $L^2$ norm in space and time, and~$F[\eta]$ maps the noise realization $\eta$ to the value of the observable $\partial_x u(0, t_0)$ for the solution of the Burgers Eq.~\eqref{eq:burgers} for the given noise realization $\eta$
, i.e.\ $F[\eta] = \partial_x u[\eta](0,t_0)$. Calculating the first-order necessary conditions for an extremum of the action in Eq.~\eqref{eq:path_integral} under the constraint $F[\eta] = a$ using the adjoint-state method~\cite{schorlepp-tong-etal:2023,schorlepp-grafke:2025}, we obtain the instanton equations, as already derived in Ref.~\cite{gurarie-migdal:1996}:
\begin{align}
    &\begin{cases}
                \partial_{t} u^{\text{I}}_{a,0} +u^{\text{I}}_{a,0}  \partial_{x} u^{\text{I}}_{a,0}-\partial_{x x} u^{\text{I}}_{a,0}=\chi * p^{\text{I}}_{a,0},  \\
                \partial_{t} p^{\text{I}}_{a,0}+u^{\text{I}}_{a,0} \partial_{x} p^{\text{I}}_{a,0}+\partial_{x x} p^{\text{I}}_{a,0}=0, 
    \end{cases}
    \label{eq:inst-eq-burgers}\\
    \text{s.t.} &\begin{cases}
         u^{\text{I}}_{a,0}(\cdot, -\infty)=0\,, \quad \partial_x u^{\text{I}}_{a,0}(0, t_0)=a\,,\\
         p^{\text{I}}_{a,0}(x, t_0)= - \lambda_a \delta'(x)
    \end{cases}
\end{align}
Here, $\lambda_a \in \mathbb{R}$ is a Lagrange multiplier, chosen such that the final-time constraint on $u$ is satisfied. We will call the solutions of these equations instantons, regardless of whether they refer to the velocity field $u^{\text{I}}$, the response field/conjugate momentum density $p^{\text{I}}$, or the corresponding noise realization $\eta^{\text{I}} = \chi^{1/2}*p^{\text{I}}$. To calculate the instantons numerically, we follow the optimal control-based approach described in Ref.~\cite{schorlepp-grafke-may-etal:2022}. Beyond the field configurations themselves, which are the most likely realizations of the fields under the constraint that a rare, large gradient strength $\partial_x u(0, t_0) = a$ is observed, this calculation gives the leading-order approximation of the one-point gradient PDF~\eqref{eq:path_integral} through the instanton action $S^{\text{I}}(a) = \tfrac{1}{2} \lVert \eta^{\text{I}}_{a,0} \rVert_{L^2}^2 = \tfrac{1}{2} \langle p^{\text{I}}_{a,0}, \chi *  p^{\text{I}}_{a,0}  \rangle_{L^2}$ via $\log \rho(a) \simeq \log \rho^{(0)}(a) := - S^{\text{I}}(a) / \sigma^2$.

Next, to approximate the one-point gradient PDF~$\rho(a)$ to 1-loop order,
we require the quadratic (Gaussian) fluctuations around the instanton to calculate the prefactor in the evaluation of Eq.~\eqref{eq:path_integral} via Laplace's method. This can be achieved by iteratively calculating the dominant eigenvalues~$\mu_a^{(k)}$ of the operator~$B_a$ in the Fredholm determinant \cite{schorlepp-tong-etal:2023,schorlepp-grafke:2025}
\begin{align}
\begin{aligned}
     C(a) &= \lvert \lambda_a \rvert \left[2 S^{\text{I}}(a) \operatorname{det}\left(\operatorname{Id}-B_a\right)\right]^{-1/2}\\
     &= \lvert \lambda_a \rvert \left[2 S^{\text{I}}(a) \prod_{k=1}^\infty \left(1-\mu_a^{(k)}\right)\right]^{-1/2}\,.
\end{aligned}
    \label{eq:prefac}
\end{align}
Here, applying the (weighted and $\left(\eta^{\text{I}}_{a,0} \right)^\perp$-projected) second variation operator $B_a$ of the solution map $F$ to a noise fluctuation $\delta \eta$ involves solving
\begin{align*}
    \left\{\begin{array}{ll}
                \partial_{t} \delta u=-\partial_{x}\left(u^{\text{I}}_{a,0} \delta u\right)+\partial_{x x} \delta u+\chi^{1 / 2} * \delta \eta, 
                & \delta u(\cdot, 0)=0, \\ 
                \partial_{t} \delta p=- \delta u \partial_{x} p^{\text{I}}_{a,0} -u^{\text{I}}_{a,0} \partial_{x} \delta p-\partial_{x x} \delta p, 
                & \delta p(\cdot, T)= 0
                \end{array}\right.
\end{align*}
and returning $\chi^{1/2} * \delta p$. For details, we refer to Refs.~\cite{schorlepp-tong-etal:2023,schorlepp-grafke:2025}.
All in all, knowing the instantons and the Gaussian fluctuations around the instanton, the one-point PDF for velocity gradients $\rho(a)$ can be expressed -- and is indeed feasible to evaluate -- to 1-loop order as
\begin{align}
\rho(a) &= \left( 2 \pi \sigma^2 \right)^{-1/2} C(a) \exp \left\{-\frac{S^{\text{I}}(a)}{\sigma^2} \right\} \left(1 + {\cal O}(\sigma^2) \right) \nonumber\\
&=: \rho^{(1)}(a)\left(1 + {\cal O}\left(\sigma^2\right) \right)
\label{eq:one-loop-grad}
\end{align}

\section{Numerical details for instanton computations and DNS}
\label{sec:num-details}

\begin{figure}
    \centering
    \includegraphics[width=\linewidth]{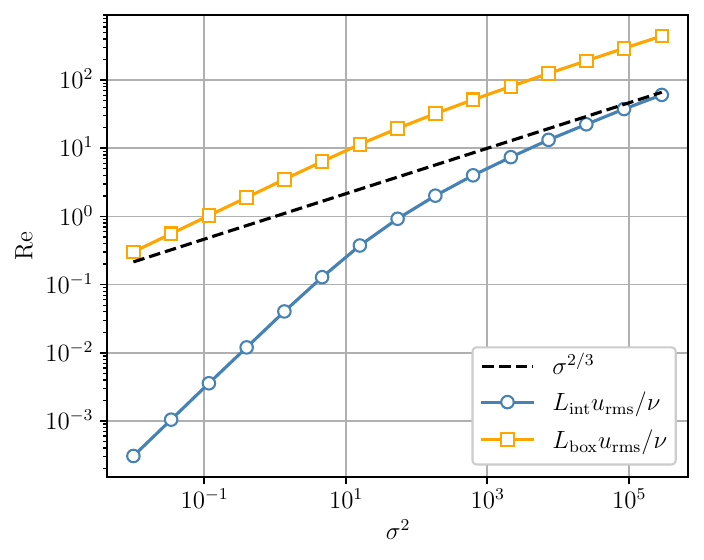}
    \caption{Reynolds number $\text{Re}$ as a function of the noise variance $\sigma^2$ for DNS of the Burgers Eq.~\eqref{eq:burgers}. The dahsed line $\text{Re} \propto \sigma^{2/3}$ corresponds to the theoretically expected scaling at large $\sigma$~\cite{grafke-grauer-schaefer-etal:2015,schorlepp-grafke-grauer:2021}. For all further results, we report $\text{Re} = L_{\text{box}} u_{\mathrm{rms}} / \nu$ (orange squares) to label the simulations.}
    \label{fig:reynolds}
\end{figure}

\begin{figure}
    \centering
    \includegraphics[width=\linewidth]{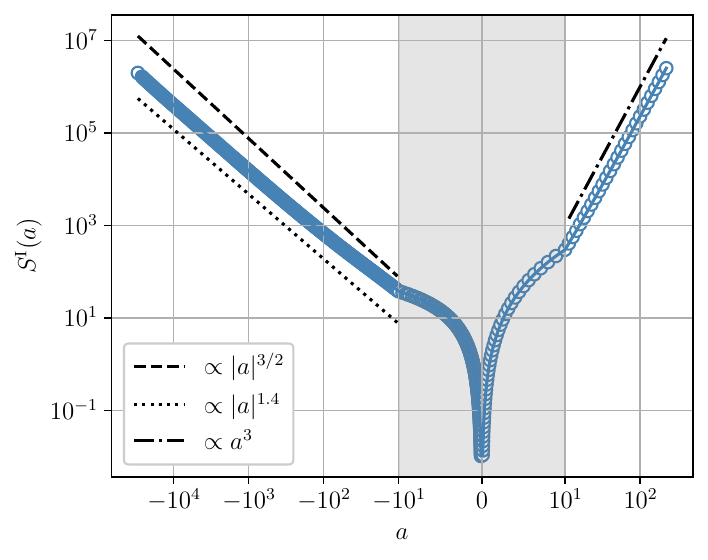}
    \includegraphics[width=\linewidth]{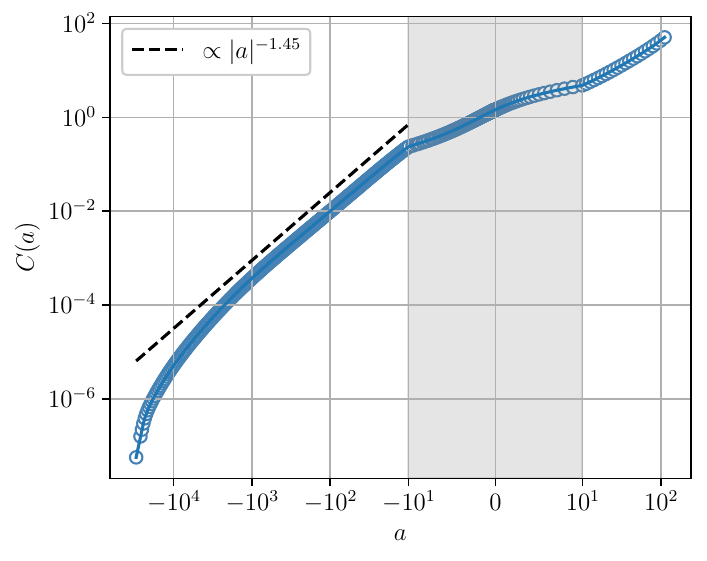}
    \caption{Numerically computed instanton action $S^{\text{I}}$ (top) and 1-loop PDF prefactor $C$ (bottom) from Eq.~\eqref{eq:one-loop-grad} as a function of the gradient $a = \partial_x u(0,t_0)$. Each open dot corresponds to one instanton configuration. Both axes are logarithmically scaled, except for the shaded region around $a=0$ where the horizontal axis is linear. Scaling lines are shown for comparison; the ones for the action at large positive and negative gradients values are in accordance with the literature~\cite{gurarie-migdal:1996,balkovsky-falkovich-kolokolov-etal:1997,chernykh-stepanov:2001,grafke-grauer-schaefer-etal:2015}.}
    \label{fig:action-prefac}
\end{figure}

As a prerequisite for all subsequent instanton-based synthetic turbulence sampling strategies, we numerically calculate and store instanton configurations $u^{\text{I}}_{a,0}(x,t)$ for about $500$ values of $a \in [-30000, -0.1]$ and about $70$ values of $a \in [0.1, 223]$, both logarithmically spaced, with an optimization code that employs a pseudo-spectral discretization and a second-order explicit Runge--Kutta time stepper with integrating factor for the diffusion terms, cf.\ Ref.~\cite{schorlepp-grafke-may-etal:2022}. We use a spatial resolution of $n_x = 512$ for smaller gradient strengths $a$ and $n_x = 2048$ for the largest ones, as well as a temporal resolution of $n_t = 10^4$ in the time interval $[-T, t_0] = [-1, 0]$ for all calculations. The size of the time interval has been chosen such that DNS statistics are approximately stationary at the final time $t_0 = 0$. The ``Mexican hat'' forcing correlation in Fourier space reads
\begin{align}
    \hat{\chi}(k) = \sqrt{2 \pi} k^2 e^{-k^2/2}\,,
    \label{eq:forcing-mexican}
\end{align}
and hence the random forcing effectively has e.g.\ only $8$ modes with an amplitude of more than $10^{-14}$. To approximate the 1-loop prefactor $C(a)$ of the one-point gradient PDF in Eq.~\eqref{eq:prefac}, we calculate the $200$ dominant eigenvalues of $B_a$ for each~$a$ as explained in Refs.~\cite{schorlepp-tong-etal:2023,schorlepp-grafke:2025}. These computations and their results, discussed below, expand upon the preliminary results using Riccati differential equations for the prefactor $C(a)$ of gradient statistics for the Burgers Eq.~\eqref{eq:burgers} shown in Ref.~\cite{schorlepp-grafke-grauer:2021}.

Similarly, we perform DNS of the Burgers Eq.~\eqref{eq:burgers} with the same physical parameters and discretization as above, for $15$ different logarithmically spaced noise strengths (and hence different Reynolds numbers) $\sigma^2 \in [10^{-2}, 3 \cdot 10^5]$ with $n_x = 2048$ and time resolutions $n_t = n_t(\sigma^2)$ chosen to ensure stability. We collect $N = 10^4$ independent samples of the velocity field at final time $u(\cdot, 0)$ for each noise strength $\sigma^2$. This constitutes the underlying ``ground truth'' ensemble of random fields that we will subsequently approximate through a gas of instantons.

\begin{figure*}
    \centering
    \includegraphics[width=.85\linewidth]{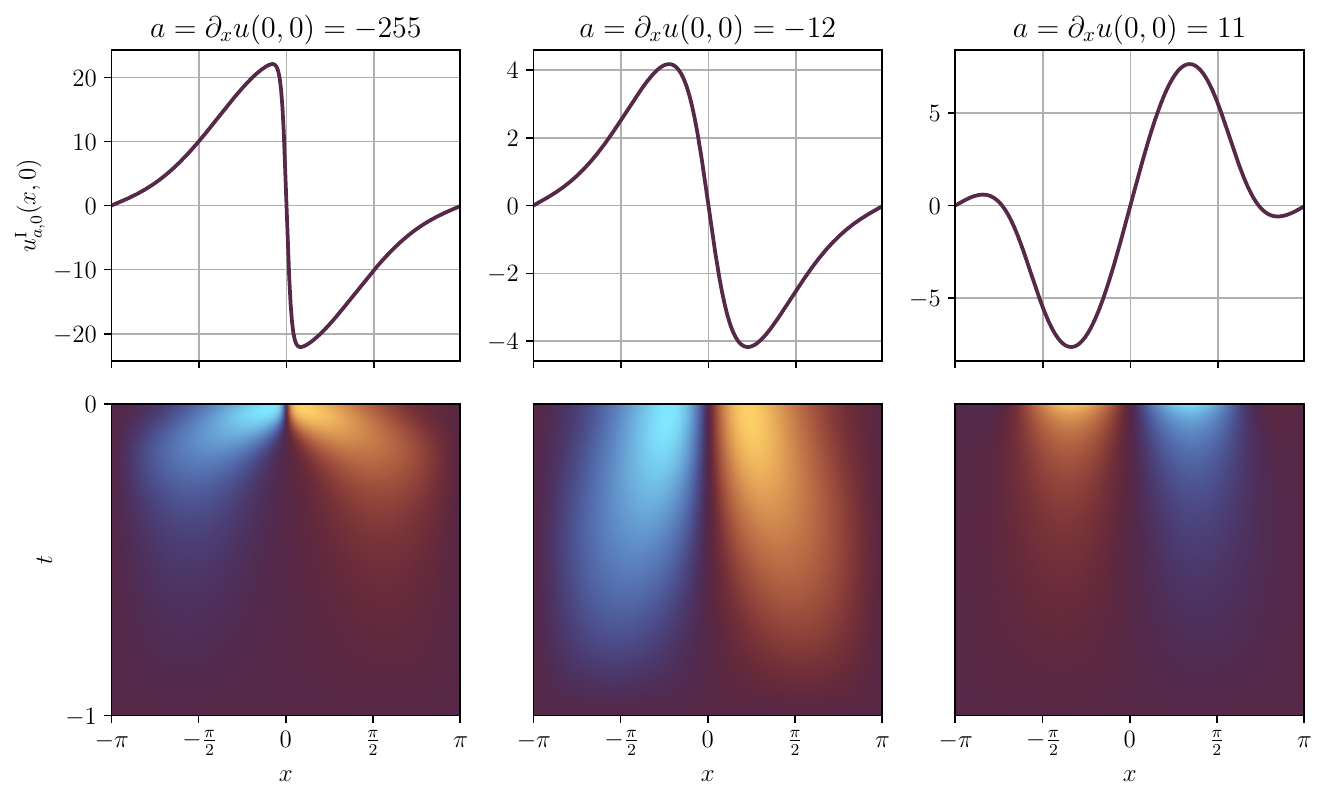}
    \caption{Numerically computed instanton configurations $u^{\text{I}}_{a,0}$ at final time $t = 0$ (top) and in space and time (bottom) for three different observable values $\partial_x u (0,0) = a \in \{-255, -12, 11\}$ (columns). For negative gradient values, we see a shock-like structure that steepens and becomes more localized in time as the absolute value of the prescribed gradient at the final time at the origin increases. For positive gradient values, we see a smoother ramp-like structure.}
    \label{fig:inst-fields}
\end{figure*}

\begin{figure*}
    \centering
    \includegraphics[width=\linewidth]{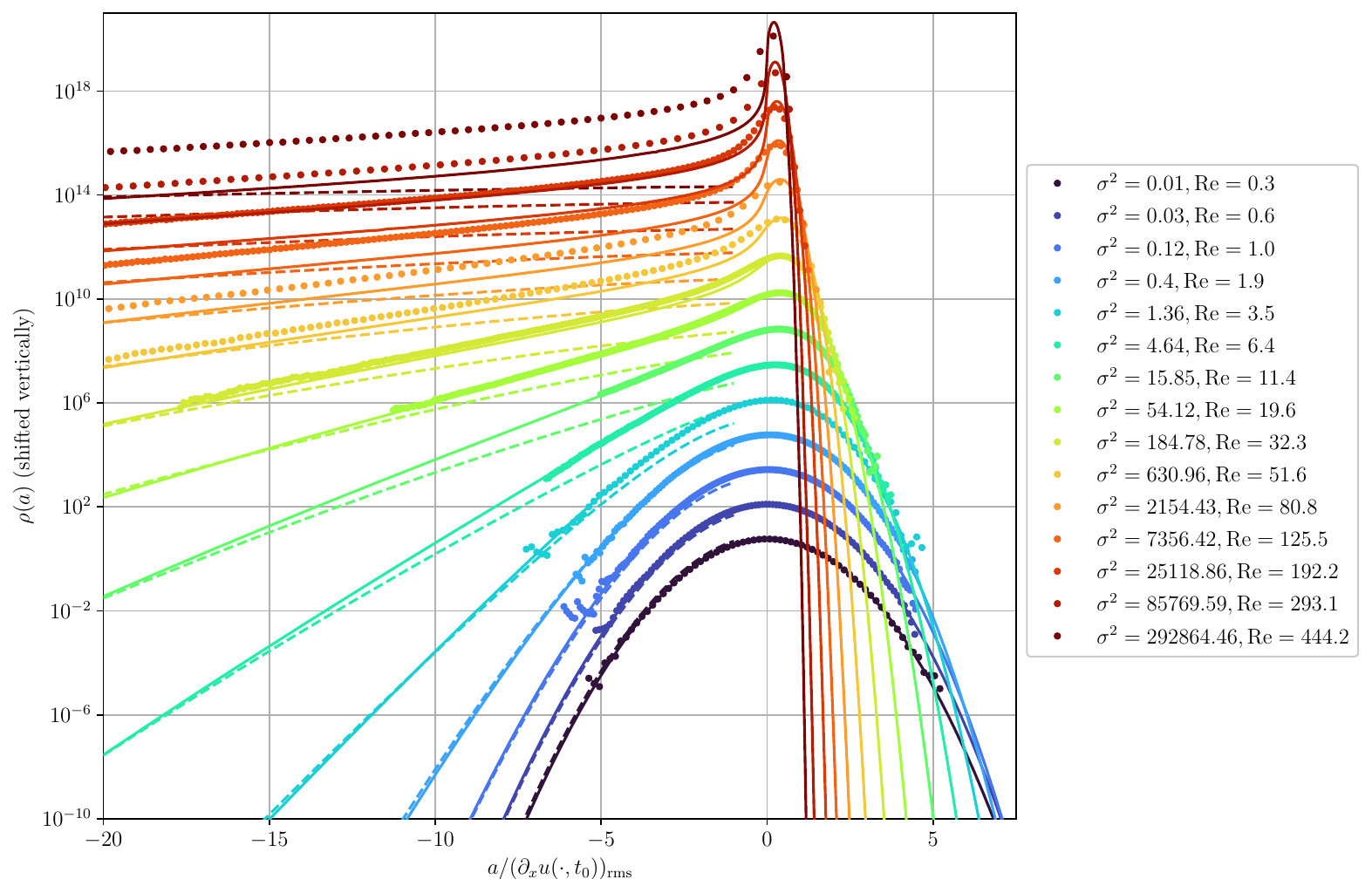}
    \caption{One-point gradient PDF $\rho$ at different Reynolds numbers from DNS of the Burgers Eq.~\eqref{eq:burgers} (dots), compared to the instanton prediction with 1-loop correction $\rho^{(1)}$ in Eq.~\eqref{eq:one-loop-grad} (solid lines). The horizontal axis has been rescaled to the respective root-mean-square gradient strength of each DNS run, and both the DNS data and instanton prediction have been shifted vertically by the same factor per Reynolds number for better visulatization. We also show the left tail of $\rho^{(0)}(a) = \text{const} \cdot \exp\left\{-S^{\text{I}}(a)/\sigma^2\right\}$ (dashed lines), i.e.\ the theoretical instanton prediction using only the instanton, with an unknown (and hence assumed to be constant) prefactor.}
    \label{fig:ux-pdf-compare-inst}
\end{figure*}

\begin{figure}
    \centering
    \includegraphics[width=\linewidth]{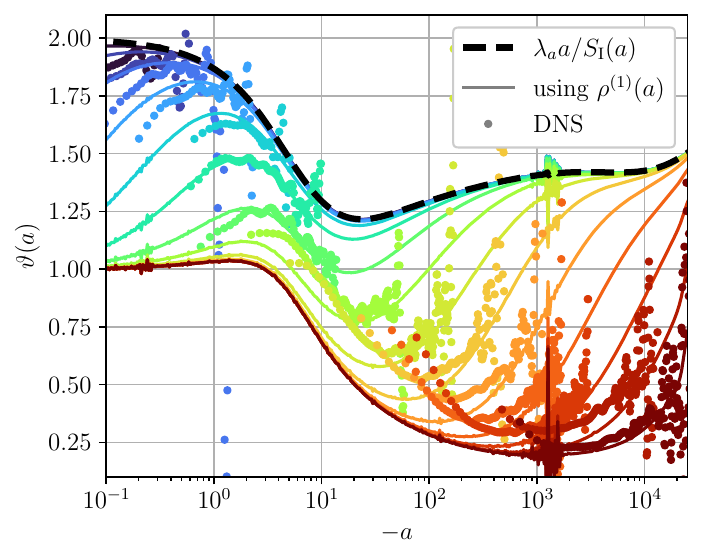}
    \caption{Local scaling exponent $\vartheta$, defined in Eq.~\eqref{eq:def-theta}, of the one-point gradient PDFs from Fig.~\ref{fig:ux-pdf-compare-inst} for negative gradients $a < 0$. The colors are the same as in Fig.~\ref{fig:ux-pdf-compare-inst}, with blue meaning small Re, and red large Re. The dashed black line is the theoretical prediction using only the instanton, while the solid lines show the (Re-dependent) predictions using the instanton and Gaussian fluctuations. For the latter, some numerical artifacts from inaccuracies in the instanton optimization and prefactor computation, amplified through the use of finite differences for $\mathrm{d}{\cal S} / \mathrm{d} a$, are visible.}
    \label{fig:theta}
\end{figure}

We show the Reynolds numbers $\text{Re}$ that corresponds to the DNS runs at different $\sigma^2$ in Fig.~\ref{fig:reynolds}. Here, we computed the root-mean-square velocity $u_{\text{rms}} = \mathbb{E
} \left[u(\cdot, t_0)^2 \right]^{1/2}$, energy $E = \tfrac12 u_{\text{rms}}^2$, the mean dissipation density $\epsilon = 2 \nu \mathbb{E} \left[ \partial_x u(\cdot, t_0)^2 \right]$, and the integral length $L_{\text{int}} = E^{3/2} / \epsilon$ from the DNS data at final time for each noise variance $\sigma^2$. We see that the DNS runs cover about three decades of $\text{Re}$. Next, we show the results of the instanton and Gaussian fluctuation computations for the necessary quantities $S^{\text{I}}(a)$ and $C(a)$ to approximate the one-point gradient PDF in Fig.~\ref{fig:action-prefac}. We remark that the power-law scaling of the prefactor $C(a)$ over a wide range of gradient strengths could be the subject of a future theoretical study. Plots of the computed shock or ramp-like instanton configurations $u^{\text{I}}_{a, 0}(x,t)$ for a few different values of $a$ are shown in Fig.~\ref{fig:inst-fields}. We refer to Ref.~\cite{chernykh-stepanov:2001} for a detailed discussion of the time evolution of instantons for the Burgers Eq.~\eqref{eq:burgers}.

By combining the instanton action~$S^{\text{I}}(a)$ and 1-loop prefactor~$C(a)$ according to Eq.~\eqref{eq:inst-sum-pdf}, we obtain the prediction for the one-point gradient PDF $\rho$ at different~$\sigma^2$ shown in Fig.~\ref{fig:ux-pdf-compare-inst} (solid lines), which compares them to the gradient PDFs as measured in the DNS runs. We see excellent quantitative agreement for small to moderate noise variances~$\sigma^2$. As expected, for larger~$\sigma^2$ the agreement becomes worse, even though the scaling of the DNS PDFs still appears to be captured correctly by the ``instanton-plus-Gaussian-fluctuations'' approximation. Importantly, as noted in Ref.~\cite{schorlepp-grafke-grauer:2021} already, we see that the 1-loop prefactor correction~$C(a)$ in~$\rho^{(1)}(a)$ is crucial for this agreement: The dashed lines~$\propto \rho^{(0)}(a)$ in Fig.~\ref{fig:ux-pdf-compare-inst} highlight that the instanton action, without accounting for fluctuations, indeed only describes the far left tail of the PDF~\cite{grafke-grauer-schaefer-etal:2015,apolinario-moriconi-pereira:2019}.\\

For a more quantitative analysis of the one-point gradient PDF scaling in Fig.~\ref{fig:ux-pdf-compare-inst}, we consider the local scaling exponent~\cite{chernykh-stepanov:2001,grafke-grauer-schaefer-etal:2015}
\begin{align}
    \vartheta(a) := \frac{\mathrm{d} \log {\cal S}(a)}{\mathrm{d} \log \lvert a \rvert} = \frac{a}{{\cal S}} \frac{\mathrm{d} {\cal S}}{\mathrm{d} a}\,.
    \label{eq:def-theta}
\end{align}
Here, we define ${\cal S}$, the exponential part of the PDF $\rho$, as follows: For either the DNS results or any of the PDF approximations through instantons at a given $\sigma^2$, we write
$\rho(a) = c \exp \{ -{\cal S}(a) / \sigma^2 \}$.
The constant $c = c(\sigma^2)$ is fixed through the condition ${\cal S}(0) \overset{!}{=} 0$~\footnote{Otherwise one could add any additive constant to ${\cal S}$ and compensate for it in $c$, rendering $\vartheta$ not uniquely defined}, such that ${\cal S}(a) = -\sigma^2 \log (\rho(a) / \rho(0))$. If the local scaling exponent $\vartheta(a)$ has a plateau somewhere, then the PDF follows a stretched exponential distribution $\propto \exp \{- \beta \lvert a \rvert^{\vartheta}\}$ with exponent~$\vartheta$ in that region. This is of particular interest, since such distributions are known to occur for extreme events in other turbulent systems as well~\cite{buaria-pumir-bodenschatz:2019}. If we approximate the PDF using just the instanton without fluctuations via~$\rho^{(0)}$, we have ${\cal S}(a) = S^{\text{I}}(a)$, and by the sensitivity property of Lagrange multipliers, the predicted exponent is $\vartheta(a) = \lambda_a a / S^{\text{I}}(a)$ for any $\sigma^2$. For the 1-loop approximation~$\rho^{(1)}$ in Eq.~\eqref{eq:one-loop-grad} including Gaussian fluctuations, we have ${\cal S}(a) = S^{\text{I}}(a) - \sigma^2 \log(C(a) / C(0))$, and hence there is a $\sigma^2$-dependent tradeoff between the action and prefactor. For both $\rho^{(1)}$ and the DNS PDFs, we use first-order finite differences to approximate the derivative in Eq.~\eqref{eq:def-theta}. Since the DNS data is noisy, in particular in the tails, we subsequently apply a smoothing filter to the resulting $\vartheta(a)$ for a cleaner visualization. The result of this procedure is shown in Fig.~\ref{fig:theta} for the more interesting case of negative gradients. The figure confirms that (i) for large $\lvert a \rvert$, the prefactor always becomes subdominant at any $\sigma^2$, and the scaling exponent hence asymptotically approaches the instanton prediction $\vartheta(a) = a \lambda_a / S^{\text{I}}(a)$ as $\lvert a \rvert \to \infty$, (ii) the instanton prediction without fluctuations interpolates non-monotonically between Gaussian scaling $\vartheta = 2$ for small~$\lvert a \rvert \ll 1$, which follows from linearizing Eq.~\eqref{eq:inst-eq-burgers}, and $\vartheta = 3/2$ asymptotically as $a \to - \infty$~\cite{balkovsky-falkovich-kolokolov-etal:1997}, and (iii) for finite $\lvert a \rvert$ and high $\sigma^2$, the local scaling exponent in the DNS data can be very far from $a \lambda_a / S^{\text{I}}(a)$, but still turns out to be well-described by the scaling exponent of $\rho^{(1)}$. Point (iii) confirms our prior observation that the scaling of the DNS PDFs is captured correctly using the instanton and Gaussian fluctuations around it, even at the highest Reynolds number considered here. The mismatch in the absolute normalization of the PDFs, which leads to the constant offset observed at high Re in Fig.~\ref{fig:ux-pdf-compare-inst}, could hence be viewed as an artifact of misrepresenting the core of the PDF at high Re.

\section{Instanton gas models}
\label{sec:inst-gas}
In this section, we give an overview of the necessary theory to formulate the probability distributions of an ensemble of superimposed single-instanton configurations. In principle, we can distinguish between the following ensembles of instantons, all of which will in general result in different statistical properties of the corresponding random fields: The instanton ensemble can be \textit{canonical}, i.e.\ composed of a fixed number $n_{\text{I}} = 1,2,3,\dots$ of instantons, or it can be \textit{grand canonical}, meaning $n_{\text{I}}$ is an integer-valued random variable that is allowed to fluctuate between different realizations drawn from the ensemble. Independently from this, we can either consider \textit{interacting} instantons, where the difference between the action of a sum of instantons and the sum of their individual actions is fully taken into account, or \textit{non-interacting} instantons as a vast simplification thereof. Between these two extremes, different models for the effective interactions among instantons can be used, e.g.\ neglecting three-body interactions and higher-order terms. Lastly, we can consider either fields composed of \textit{only instantons}, or furthermore add -- as a first and practically accessible approximation -- random \textit{Gaussian fluctuations} around the individual instantons as well. We will discuss different combinations of these properties in the following. Afterwards, Sec.~\ref{sec:mc} then focuses on how to actually sample from the distributions introduced in this section in practice.

\subsection{Canonical ensemble of instantons}
\label{sec:canonical}
The most straightforward instanton gas ansatz we can take for the random field $u = u(x,t)$ is, for a fixed and prescribed number of instantons $n_{\text{I}}$, a sum of instantons without any random fluctuations around them~\footnote{It is in fact not crucial for the basic approach of this section that the superimposed fields $u = u^{\text{I}}_{a,x_0}$ are instanton solutions with $\delta S / \delta u = 0$; the target distribution~\eqref{eq:inst-sum-pdf} would remain formally unchanged for any parametric family of fields that have zero spatio-temporal mean, cf.\ the discussion in Ref.~\cite{ivashkevich:1997}.}
\begin{align}
    u  = \sum_{j = 1}^{n_{\text{I}}} u^{\text{I}}_{a_j,x_{0j}} = u\left( \left\{a_j,x_{0j} \right\}_{j=1}^{n_{\text{I}}} \right)\,.
    \label{eq:u-sum-inst}
\end{align}
This corresponds to a canonical ensemble in the language of statistical mechanics. Consequently, the free and random parameters of this instanton gas model are only the positions $x_{0j} \in [0, L_{\text{box}})$ and gradient strengths $a_j \in \mathbb{R}$ of the individual instantons $j = 1, \dots, n_{\text{I}}$. This parameterizes a finite-dimensional submanifold in the space of all field realizations. The PDF of the original ensemble in the space of all fields, corresponding to DNS of the Burgers Eq.~\eqref{eq:burgers}, is given by
\begin{align}
    \rho[u] \propto \exp \left\{-\frac{1}{\sigma^2} S[u] \right\}
\end{align}
with classical Onsager--Machlup~\cite{onsager-machlup:1953,machlup-onsager:1953} or Freidlin--Wentzell~\cite{freidlin-2012} action functional
\begin{align}
    S[u] = \frac{1}{2} \left \langle \partial_t u + u \partial_x u - \partial_{xx}u,  \chi^{-1} * [\dots] \right \rangle_{L^2}\,,
    \label{eq:om-action}
\end{align}
where $[\dots]$ stands for repeating the preceding expression. Here, the additional Jacobian term from the noise-to-velocity transformation~\cite{zinn-justin:2021,gladrow-keyser-adhikari-etal:2021} in the action at finite~$\sigma$, proportional to the functional divergence of the deterministic terms in the equation of motion $u \partial_x u - \partial_{xx}u$, can be neglected~\cite{ivashkevich:1997}, e.g.\ because the instantons for the Burgers Eq.~\eqref{eq:burgers} are all antisymmetric functions in space with respect to $x_0$. Now, restricting the fields to lie on the submanifold described by Eq.~\eqref{eq:u-sum-inst} leads to
\begin{align}
    \rho \left( \left\{a_j,x_{0j} \right\}_{j=1}^{n_{\text{I}}} \right) &\propto g\left( \left\{a_j,x_{0j} \right\}_{j=1}^{n_{\text{I}}} \right) \times \nonumber\\
    &\hspace{-.5cm} \times\exp \left\{-\frac{1}{\sigma^2} S\left[u\left( \left\{a_j,x_{0j} \right\}_{j=1}^{n_{\text{I}}} \right)\right] \right\}
    \label{eq:inst-sum-pdf}
\end{align}
as the induced PDF of the instanton coordinates $\left\{a_j,x_{0j} \right\}_{j=1}^{n_{\text{I}}}$, up to a normalization constant.
Here, $g$ denotes the Gram determinant of the hyper-surface parameterized by Eq.~\eqref{eq:u-sum-inst}. Concretely, because of the sum ansatz and translation invariance, the Gramian factorizes into
\begin{align}
    g \left( \left\{a_j,x_{0j} \right\}_{j=1}^{n_{\text{I}}} \right) = \prod_{j = 1}^{n_{\text{I}}} g \left( a_j \right)
\end{align}
with
\begin{align}
    g\left(a\right) =  \bigg[ &\left \lVert \partial_{a} u^{\text{I}}_{a,0} \right \rVert_{L^2}^2 \left \lVert \partial_{x} u^{\text{I}}_{a,0} \right \rVert_{L^2}^2 - \left \langle \partial_{a} u^{\text{I}}_{a,0}, \partial_{x} u^{\text{I}}_{a,0} \right \rangle_{L^2}^2 \bigg]^{1/2}
    \label{eq:gramian-result}
\end{align}
in this case. Note that the Gramian $g$, despite appearing as a pre-exponential factor in the PDF~\eqref{eq:inst-sum-pdf}, is purely geometric in nature, and does not correspond to any random fluctuations around the instanton fields.\\

Adding and subtracting the single-instanton actions in the exponent of Eq.~\eqref{eq:inst-sum-pdf}, we can alternatively write
\begin{align}
    \rho \left( \left\{a_j,x_{0j} \right\}_{j=1}^{n_{\text{I}}} \right) &\propto \left( \prod_{j=1}^{n_{\text{I}}} g(a_j) \rho^{(0)}(a_j) \right) \times \nonumber\\
    &\times\exp \left\{-\frac{1}{\sigma^2} \Delta S\left( \left\{a_j,x_{0j} \right\}_{j=1}^{n_{\text{I}}} \right) \right\}
    \label{eq:multi-inst-pdf}
\end{align}
    where the \textit{interaction term}, given by the action difference, is defined as
\begin{align}
    \Delta S\left( \left\{a_j,x_{0j} \right\}_{j=1}^{n_{\text{I}}} \right) =  S\left[ \sum_{j= 1}^{n_{\text{I}}} u^\text{I}_{a_j, x_{0j}} \right] -  \sum_{j = 1}^{n_{\text{I}}} S \left[ u^{\text{I}}_{a_j, x_{0j}} \right]\,,
    \label{eq:delta-s}
\end{align}
and we recall that $\rho^{(0)}(a) = \exp\{-S [ u^{\text{I}}_{a,0} ] / \sigma^2 \}$ is the leading-order approximation of the one-point gradient PDF $\rho(a)$ using just the instanton.

The interaction $\Delta S$ in Eq.~\eqref{eq:delta-s} is a function of all instanton coordinates $a_j$ and $x_{0j}$ in general; we would for example expect that it tends to zero for well-separated (in terms of the $x_0$'s) instantons. For a \textit{non-interacting} gas of instantons, we would simply approximate $\Delta S \equiv 0$ for any configuration of $n_{\text{I}}$ instantons; then the PDF~\eqref{eq:multi-inst-pdf} factorizes into a product
\begin{align}
    \rho \left( \left\{a_j,x_{0j} \right\}_{j=1}^{n_{\text{I}}} \right) \propto \prod_{j=1}^{n_{\text{I}}} g(a_j) \rho^{(0)}(a_j)
    \label{eq:pdf-noninter}
\end{align}
of individual PDFs that can easily be sampled from, see the next section. As mentioned above, as an intermediate step between setting $\Delta S \equiv 0$ on the one hand, and exactly evaluating $\Delta S$ according to Eq.~\eqref{eq:delta-s} on the other hand, there is also a whole range of effective interaction models we can use to simplify the PDF and render sampling from it more efficient, e.g.\ taking into account only two-body interactions
\begin{align}
    \Delta S\left( \left\{a_j,x_{0j} \right\}_{j=1}^{n_{\text{I}}} \right) \approx \tfrac{1}{2} \sum_{j \neq k} U^{(2)}\left(a_j, x_{0j}, a_k, x_{0k} \right)
\end{align}
of two instantons, where an ansatz, or systematic approximation based on limits of Eq.~\eqref{eq:delta-s}, for the function~$U^{(2)}$ directly in terms of the parameters $a_j, x_{0j}, a_k, x_{0k}$ is used, cf.\ Refs.~\cite{schaefer-shuryak:1996,munster-kamp:2000}. The main advantage is that with such an interaction model, Monte Carlo simulations or analytical calculations can be performed only in terms of the collective coordinates $\left\{a_j,x_{0j} \right\}_{j=1}^{n_{\text{I}}}$, and it is not necessary to assemble and calculate the action of the full space-time dependent fields $u(x,t)$. In this paper, we will only consider the case of fully accounting for interactions via Eq.~\eqref{eq:delta-s} for simplicity, but a closer study of effective interactions between instantons in fluid dynamics would be an interesting future work. 

Finally, we note that our approach in this section is similar to Ref.~\cite{millo-faccioli-scorzato:2010}, who further integrate over all random fluctuations locally $L^2$-orthogonal to the instanton-sum to determine the effective interaction term, for example through Monte Carlo simulations or perturbation theory. In this sense, Eq.~\eqref{eq:multi-inst-pdf} with $\Delta S$ given by Eq.~\eqref{eq:delta-s} can be understood as a zeroth-order approximation thereof, neglecting noise corrections.

\subsection{Grand canonical ensemble of instantons and the choice of the number of instantons $n_{\text{I}}$}
\label{sec:grand-canonical}

It is intuitive to expect that if -- instead of prescribing a fixed number of instantons $n_{\text{I}}$ in Eq.~\eqref{eq:u-sum-inst} -- the number of instantons $n_{\text{I}}$ is allowed to vary, with a self-consistently determined distribution, then a better approximation to the true random field ensemble could be obtained. This corresponds to using a grand canonical ensemble of instantons in the language of statistical mechanics. If fluctuations in the number of instantons around $\mathbb{E} \left[n_{\text{I}} \right]$ are not negligible, we may expect it to possibly behave differently from a canonical ensemble. The relative weight and hence probability of an $n_{\text{I}}$-instanton configuration is determined by the $n_{\text{I}}$-instanton partition function. Typically, interacting particles in a gas within the grand canonical ensemble are simulated by adding random particle addition and removal moves to the Metropolis sampler of Sec.~\ref{sec:mc}~\cite{frenkel-smit:2023,ulybyshev-winterowd-assaad-etal:2023}.\\

For the present work, we leave this is a future direction to explore, and only consider canonical ensembles, with an instanton number $n_{\text{I}} = 3$ throughout, chosen heuristically in line with DNS snapshots (see the discussion below). However, we remark that in the interacting canonical instanton ensemble of the previous subsection, since an instanton $u^{\text{I}}_{a_j, x_{0j}}$ effectively ``disappears'' from Eq.~\eqref{eq:u-sum-inst} as $a_j \to 0$, it is possible that due to exclusion-type interactions, the ensemble assigns a non-negligible probability to configurations with effectively fewer than~$n_{\text{I}}$ instantons. One should hence think of the interacting canonical ensemble considered here as prescribing an \textit{upper bound} $n_{\text{I}}$ on the number of instantons, but potentially resulting in fewer instantons being present with high probability. In contrast to this, the non-interacting canonical ensemble~\eqref{eq:pdf-noninter} strictly fixes the number of instantons to $n_{\text{I}}$. In the latter ensemble, a smaller effective number of instantons in a given random field sample can e.g.\ result from independently drawing two instanton positions that are close to the same location, and with comparable absolute value of their gradient strengths of either the same or opposite sign.\\

We can estimate the typical spatial extent of a single coherent structure for the Burgers Eq.~\eqref{eq:burgers} and the corresponding instanton (not only the shock width itself, but the entire structure) as the correlation length~$L_{\chi}$ of the large-scale forcing covariance $\chi$ (in the present setting: $L_\chi = 1$, $L_\text{box} = 2 \pi$). Hence, the number of instantons~$n_{\text{I}}$ should be chosen as $L_{\text{box}} / L_{\chi}$ up to a dimensionless constant of $O(1)$. Choosing a canonical ensemble of instantons with $n_{\text{I}} \ll L_{\text{box}} / L_{\chi}$ will lead to a dilute, weakly interacting gas, whereas $n_{\text{I}} \gg L_{\text{box}} / L_{\chi}$ leads to strong interaction effects. In general, a theoretical determination of $n_{\text{I}}$ for the canonical ensembles would involve calculating or approximating $\mathbb{E}[n_{\text{I}}]$ within a grand-canonical one~\cite{schaefer-shuryak:1996}. In contrast to this, for this paper, in order to determine $n_{\text{I}}$ for numerical simulations of canonical ensembles, we counted peaks in $\partial_x u$ snapshots from DNS at different Re, which indicated that no more than $n_{\text{I}} = 4$ instantons should be used. We then repeated the simulations and (Eulerian) statistical analysis shown in Sec.~\ref{sec:results} for $n_{\text{I}} = 1, \dots, 4$, and present results for the representative value $n_{\text{I}} = 3$ for all Re in this paper, which gave the closest overall visual and statistical agreement with DNS.

\subsection{Including Gaussian fluctuations}
\label{sec:gaussian-fluc}

As a simple heuristic to possibly improve the instanton gas approximation and include smaller-scale random fluctuations around the instantons, we propose to add, at the end of the respective sampling procedures and independently from all other random variables $a_j, x_{0j}$ and potentially $n_{\text{I}}$, Gaussian fluctuations around each instanton. Their covariance will be determined according to the second variation of the action around each individual instanton. Note that these fluctuations are also what resulted in the 1-loop prefactor~$C(a)$ for the gradient PDF in Eq.~\eqref{eq:one-loop-grad} in Sec.~\ref{sec:inst-intro}.

Explicitly, by diagonalizing the second variation of the action around each instanton following Ref.~\cite{schorlepp-tong-etal:2023}, we set
\begin{align}
    u(x) = \sum_{j = 1}^{n_{\text{I}}} \left( u^{\text{I}}_{a_j,x_{0j}}(x,t_0) + \sigma \sum_{k = 1}^\infty Z_{jk} \sqrt{\nu_{a}^{(k)}} \delta v_{a,x_0}^{(k)}(x) \right)
    \label{eq:gauss-flucs}
\end{align}
with random normal coefficients $Z_{jk} \overset{\text{iid}}{\sim} {\cal N}(0,1)$ within a Karhunen--Lo{\`e}ve expansion~\cite{alexanderian:2015} of the conditioned Gaussian fluctuations at time $t_0$. Here,~$\nu_{a}^{(k)} \geq 0$ and~$\delta v_{a,x_0}^{(k)}$ are the eigenvalues and orthonormal eigenfunctions, respectively, of the integral operator $K_{a, x_0}$ acting as $(K_{a, x_0}v)(x) = \int_0^{2 \pi} {\cal C}_{a, x_0}(x,x') v(x') \mathrm{d} x'$, which is defined through the final-time covariance function
\begin{align}
{\cal C}_{a, x_0}(x,x') = \sum_{k = 1}^\infty \frac{\delta u_{a,x_{0}}^{(k)}(x,t_0) \delta u_{a,x_{0}}^{(k)}(x',t_0)}{1 - \mu_{a}^{(k)}}\,,
\end{align}
of the Gaussian fluctuations around an instanton at position~$x_0$ with gradient strength~$a$. To compute this covariance function, besides the eigenvalues~$\mu_{a}^{(k)}$ of the second variation~$B_a$ of the noise-to-observable map $F$ from Eq.~\eqref{eq:prefac}, we require the functions $\delta u =\delta u_{a,x_0}^{(k)}$ which solve the linearized Burgers equation
\begin{align}
    \partial_t \delta u+ \partial_x\left(u^{\text{I}} \cdot \delta u \right) - \partial_{xx}  \delta u = \chi^{1/2} * \delta \eta
\end{align}
along the instanton $u^{\text{I}} = u^{\text{I}}_{a, x_0}$ driven by the $k$-th orthonormal eigenfunction~$\delta \eta = \delta \eta_{a,x_0}^{(k)}$ of~$B_a$, cf.\ Sec.~\ref{sec:inst-intro}.

The idea of this approach is also discussed and implemented in Ref.~\cite{schaefer:2004} in the context of quantum mechanical tunneling. While it is somewhat self-consistent in the sense of using the best Gaussian approximation around each individual instanton in the small noise $\sigma^2 \to 0$ or extreme event $\lvert a \rvert \to \infty$ limit, it does not in fact employ the best Gaussian approximation around the sum of instantons in general. Nevertheless, as the eigenvalues~$\nu_{a}^{(k)}$ and eigenfunctions~$\delta v^{(k)}_{a, 0}$ can be precomputed along with the instantons themselves, it is easy to sample these Gaussian fluctuations in practice.

We further note that this approach may lead us to consider using the 1-loop gradient PDF~$\rho^{(1)}$ as defined in Eq.~\eqref{eq:one-loop-grad}, instead of~$\rho^{(0)}$, for the instanton parameter distribution in Eq.~\eqref{eq:multi-inst-pdf}, as is commonly done in the literature~\cite{schaefer-shuryak:1996,schaefer-shuryak:1998,schaefer:2004,ulybyshev-winterowd-assaad-etal:2023}. We will explore the consequences of these modifications -- including Gaussian fluctuations according to Eq.~\eqref{eq:gauss-flucs}, and exchanging~$\rho^{(0)}$ for~$\rho^{(1)}$ in Eq.~\eqref{eq:multi-inst-pdf} -- for turbulence statistics and particle propagation in Sec.~\ref{sec:results}. 

\section{Monte Carlo simulations of the instanton gas}
\label{sec:mc}

\begin{figure}
    \centering
    \includegraphics[width = \linewidth]{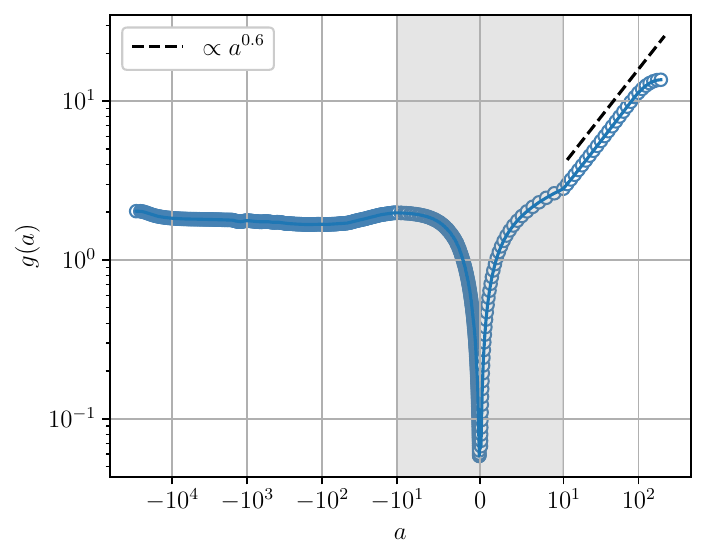}
    \caption{Gramian $g$, defined in Eq.~\eqref{eq:gramian-result}, as a function of the gradient value $a = \partial_x u(0,t_0)$, for the transformation onto the collective coordinates $(a,x_0)$ of the instantons. Each open dot corresponds to one numerically computed instanton configuration. Note the double-log scale of the figure; in the shaded region around $a=0$, a linear horizontal scale is used instead. For large negative gradient values, $g$ is almost constant.}
    \label{fig:gram}
\end{figure}

For a prescribed number $n_{\text{I}}$ of non-interacting instantons and noise strength or Reynolds number $\sigma^2$, sampling from the target density~\eqref{eq:pdf-noninter} is easily achieved by uniformly sampling instanton positions
\begin{align}
    x_{0j} \overset{\text{iid}}{\sim} {\cal U} \left([0, L_{\text{box}}] \right)\,,
\end{align}
and gradient strengths from
\begin{align}
    a_j \overset{\text{iid}}{\sim} g \cdot \rho^{(0)}\,,
    \label{eq:sample-grad}
\end{align}
all independently from each other. We calculate and store the Gramian $g(a)$ as defined in Eq.~\eqref{eq:gramian-result} in advance from the pre-computed instanton data, cf.\ Sec.~\ref{sec:num-details}. For~$\partial_a u_{a,0}^{\text{I}}$ in Eq.~\eqref{eq:gramian-result}, we simply use finite differences in~$a$ (and a subsequent smoothing operation by convolving the curve with a smooth kernel, to remove numerical artifacts resulting from the nonuniform $a$-spacing described in Sec.~\ref{sec:num-details}). The resulting function~$g$ is shown in Fig.~\ref{fig:gram}, and we observe that it is roughly constant in the left tail, and will hence have a negligible influence for high Reynolds numbers in the present problem. We note that no costly MCMC method or rejection steps are necessary to sample the non-interacting instanton ensemble. Gaussian fluctuations around individual instantons may be included \textit{a posteriori} as described in Sec.~\ref{sec:gaussian-fluc}, by simply replacing $\rho^{(0)}$ by $\rho^{(1)}$ in Eq.~\eqref{eq:sample-grad} for the one-point gradient PDF, and adding to each instanton the random perturbations from Eq.~\eqref{eq:gauss-flucs}.

For the case of interacting instantons, with joint PDF~\eqref{eq:inst-sum-pdf} for $2 n_{\text{I}}$ random parameters (instanton positions and gradients), the density no longer factorizes. Hence one needs to use e.g.\ a MCMC based approach to sample from the target density~\eqref{eq:inst-sum-pdf}, which is known up to a normalization constant. For instance, one could use the classical Metropolis algorithm~\cite{metropolis-rosenbluth-rosenbluth-etal:1953}, where in each step of the chain, the proposal could consist of
\begin{enumerate}
    \item picking an index $\hat{j} \sim {\cal U}(\{1, \dots, n_{\text{I}}\})$,
    \item changing the position $x_{0\hat{j}} \to x_{0\hat{j}}' = x_{0\hat{j}} + \delta_{x_0} \cdot Z_1$ and gradient strength  $a_{\hat{j}} \to a_{\hat{j}}' = a_{\hat{j}} + \delta_a \cdot Z_2$ with $Z_1, Z_2  \overset{\text{iid}}{\sim} {\cal N}(0,1)$,
    \item keeping the other positions and gradient strengths the same, $ x_{0k}'=x_{0k} $, $a_k' = a_k$ for all $k \neq \hat{j}$
\end{enumerate}
Here, the choice of $\delta_{x_0}, \delta_a > 0$, or other proposals, may require fine-tuning. The proposed update of the chain's state is accepted with probability $p = \min \{1,r\}$ where
\begin{align}
    &r =  \frac{ \rho \left( \left\{a_j',x_{0j}' \right\}_{j=1}^{n_{\text{I}}} \right)}{ \rho \left( \left\{a_j,x_{0j} \right\}_{j=1}^{n_{\text{I}}} \right)}
    = \frac{g\left(a_{\hat{j}}' \right)}{g\left(a_{\hat{j}} \right)} \times \\
    & \times \exp \left\{-\frac{1}{\sigma^2} \left(S\left( \sum_{j = 1}^{n_{\text{I}}} u^{\text{I}}_{a_j', x_{0j}'} \right) -  S\left( \sum_{j = 1}^{n_{\text{I}}} u^{\text{I}}_{a_j, x_{0j}} \right) \right) \right\} \,. \nonumber
\end{align}
In addition to the increased cost of running the Markov chain in itself, we see that it is necessary to calculate the action~\eqref{eq:om-action} of the full space-time configuration $u(x,t)$ in each step to compute the acceptance probability. This makes the approach, without further approximations or interaction models as briefly discussed in Sec.~\ref{sec:canonical}, computationally expensive. In particular, this would be relevant in more than one spatial dimension, and it might even be less efficient than running DNS of Eq.~\eqref{eq:burgers}. For the present work, we consider it mainly of theoretical interest to investigate how interactions between instantons influence the Eulerian and Lagrangian field statistics, and leave efficiency considerations as future work.

For this paper, for simplicity -- as it requires less fine-tuning, and a robust and parallelized implementation is already available -- we use the affine-invariant ensemble sampler of Ref.~\cite{goodman-weare:2010} as implemented by Ref.~\cite{foreman-mackey-hogg-lang-etal:2013}, instead of the Metropolis sampler, to sample the $6$-dimensional distribution~\eqref{eq:inst-sum-pdf} of the collective coordinates (for $n_{\text{I}} = 3$). For each $\sigma^2$, we run $N_{\text{w}} =  90$ interacting ``walkers'', with independent initial states $ \left\{a_j^{(k)},x_{0j}^{(k)} \right\}_{j=1}^{n_{\text{I}}}$, $k = 1, \dots, N_{\text{w}}$, sampled from the non-interacting distribution~\eqref{eq:pdf-noninter}, for $5000$ MCMC steps and measure the coordinate-wise autocorrelation time $\tau$ of the chain. A burn-in period of~$2 \tau$ is then used, and approximately independent samples (using all walkers) are subsequently obtained by subsampling the chains every~$\tau / 2$ steps. Since for the problem at hand, the velocity field~\eqref{eq:u-sum-inst} is $2 \pi$-periodic in space and the PDF~\eqref{eq:inst-sum-pdf} hence only depends on $x_{0j} \text{ mod } 2 \pi$, we shift the instanton positions back to $[0, 2 \pi]$ every $100$ MCMC steps to avoid numerical errors arising from possibly large values of the~$x_{0j}$'s.

A noteworthy technical point here is the presence of the inverse forcing correlation $\chi^{-1}$ defined by $\chi * \chi^{-1} = \delta$ in the Onsager--Machlup action~\eqref{eq:om-action}. This amounts to taking $1/\hat{\chi}(k)$ in Fourier space, which becomes singular for high $k$ modes of the large-scale forcing correlation~\eqref{eq:forcing-mexican} we used. In principle, this property correctly enforces that only fields $u$ that can actually be realized as solutions of the Burgers Eq.~\eqref{eq:burgers} under the forcing~\eqref{eq:forcing-mexican} have nonzero probability, but it can lead to numerical issues when evaluating Eq.~\eqref{eq:om-action}. Hence, when evaluating the (logarithm of the unnormalized) PDF~\eqref{eq:inst-sum-pdf} for the MCMC steps, we only take into account modes with $\hat{\chi}(k) > 10^{-5}$ for the evaluation of the action~\eqref{eq:om-action} in Fourier space via the Plancherel theorem. In this regard, and also more generally, exploring a less smooth forcing, such as power-laws $\hat{\chi}(k) \propto k^{-\alpha}$ for different exponents~$\alpha$ (cf.\ Ref.~\cite{bec-khanin:2007}), could be a valuable future direction.

Lastly, we would like to stress once again that all sampling procedures described in this and the previous section do not require \textit{any} input from DNS of the underlying equation of motion~\eqref{eq:burgers}, besides perhaps the choice of observable and $n_{\text{I}}$. Hence, they can be regarded as a systematic and first-principles approach to sampling synthetic turbulent fields within the chosen ansatz. It is also possible to combine this first-principles approach with information and data from DNS or -- in the case of magnetic turbulence -- observations from satellite missions. For example, the gradient strength from DNS~\cite{zhdankin-uzdensky-etal:2013} or the statistics of current sheets from observations~\cite{vordanova-etal:2020} could be incorporated directly.

\section{Further random field ensembles for comparison}
\label{sec:other-ensembles}

In order to assess the influence of coherent structures -- that are present in the instanton gas by definition -- on the resulting turbulence statistics more comprehensively, we will also compare our results to two other random field ensembles without structures in Sec.~\ref{sec:results}. 
We briefly introduce those two ensembles in this section.

\subsection{Phase-randomized DNS fields}
\label{sec:phase-random}

To generate random fields with the same energy spectrum as those obtained from DNS (see Sec.~\ref{sec:num-details}) of the Burgers Eq.~\eqref{eq:burgers}, but without the coherent structures, i.e.\ smooth ramps and shocks, we multiply each Fourier mode of the DNS fields with an i.i.d.\ random phase $\phi_k \sim {\cal U}([0, 2 \pi))$. This is a commonly used strategy in the literature, cf.\ Refs.~\cite{koga-chian-miranda-etal:2007,grauer-homann-pinton:2012,shukurov-snodin-seta-etal:2017} and others. Since coherent structures such as shocks require phase coherence between different Fourier modes, randomizing the phases in this way eliminates structures, intermittency and anomalous scaling. The individual fields and their gradients are then Gaussian random variables. Concretely, if $\hat{u}_k$ for $k \in \mathbb{Z}$ are the Fourier modes of the Burgers DNS fields, we consider new fields $v$ such that
\begin{align}
    \hat{v}_k = \hat{u}_k \exp \left\{i \phi_k \right\}\,, \quad k = 0, 1, 2, \dots
\end{align}
and set $\phi_{-k} = -\phi_{k}$ for the negative wave number to keep the fields real in real space, requiring $\hat{v}_{-k} = \hat{v}_{k}^*$. Clearly, this phase-randomization leaves the energy spectrum $E(k) = \mathbb{E}  \lvert \hat{u}_k \rvert^2/2 = \mathbb{E} \lvert \hat{v}_k \rvert^2/2 $ invariant.

\subsection{Lognormal cascade model}
\label{sec:lognormal-cascade}

To generate random fields with anomalous scaling but without the coherent structures, we consider the continuous wavelet cascade model introduced by Muzy~\cite{muzy:2019}.
The goal of this model, as a representative member of a class of synthetic turbulence models, is to mimic the multifractal statistical properties of turbulence, independently from the fluid's equations of motion.
The corresponding stochastic process~$u$ from Ref.~\cite{muzy:2019} is defined as the weak limit, when the small-scale cutoff $\ell$ tends to $0$, of
\begin{align}
    u_\ell(x) = \int_{\ell}^{L_{\text{int}}} \mathrm{d} s \; s^{H-2} \int_{-\infty}^\infty \mathrm{d} b\; \exp\{\omega_s(b)\} \psi\left(\tfrac{x - b}{s} \right)\,.
    \label{eq:muzy}
\end{align}
Here, $0<H<2$ is a hyperparameter that controls the smoothness of realizations (similar to the Hurst exponent for fractional Brownian motion), with smaller $H$ corresponding to rougher realizations.
Randomness is introduced through the stochastic process~$\omega_s(x)$, which is infinitely divisible in the scale~$s$.
The non-linear operation~$\exp\{\omega_s(x)\}$ introduces intermittency and fulfills the multifractal scaling relation $\mathbb{E}[\exp\{p\,\omega_s(x)\}]=\left(L_\mathrm{int}/s\right)^{\phi(p)}$, where $\phi(p)$ denotes the cumulant generating function of $\omega_s(x)$.
This construction constitutes the continuous counterpart to classical discrete multiplicative cascades~\cite{frisch:1995}.
The multifractal noise~$\exp\{\omega_s(x)\}$ is translated scale-by-scale into the velocity signal~$u(x)$ in Eq.~\eqref{eq:muzy} via convolution with the synthesizing wavelet function~$\psi(x)$.

We focus here on the simple log-normal cascade given by a Gaussian process~$\omega_s(x)$ with the cumulant generating function $\phi(p)=\frac{\mu}{2}p(p-1)$, where the hyperparameter~$\mu>0$ controls the intensity of the intermittency.
The structure function scaling exponents of the velocity signal~$u(x)$ (see Eq.~\eqref{eq:struc-func} below for the definition) are hence expected to follow
\begin{equation}
    \zeta_p=pH-\frac{\mu}{2}p(p-1)\,.
    \label{eq:zetap}
\end{equation}

Our objective is now to match the energy spectra and structure function scaling exponents of the Burgers data as close as possible in order to assess the role of multifractality without coherent structures.
First, we choose the Haar wavelet
\begin{equation}
    \psi(x)=\begin{cases}
        1 &\quad 0\le x<\frac{1}{2}, \\
        -1 &\quad \frac{1}{2}\le x<1, \\
        0 &\quad\text{otherwise},
    \end{cases}
\end{equation}
as the synthesizing function, because it adequately resembles the viscous cutoff of the Burgers fields at high wavenumbers.
The finite cutoff wavenumber~$k_\mathrm{cutoff}$ is then fitted to the data and numerically linked to a finite cutoff scale~$\ell$ via~$\ell\approx0.57/k_\mathrm{cutoff}$ assuming the energy spectrum decays asymptotically as~$\sim\exp(-(k/k_\mathrm{cutoff})^{1.5})$.
As the noise strength~$\sigma^2$ increases, the cutoff scale~$\ell$ tends to zero.
The integral scale is set to~$L_\mathrm{int}=2\pi$.

Next, we approximate the low-order Burgers scaling~$\zeta_p^\text{Burgers}$ with log-normal scaling~$\zeta_p^\text{log-normal}$ by fitting Eq.~(\ref{eq:zetap}) to the numerically obtained~$\zeta_p^\text{Burgers}$ for~$p<5$.
We denote the thusly obtained parameters as~$H_\text{actual}$ and~$\mu_\text{actual}$.
When generating sample signals~$u(x)$ with these parameters, we notice that Eqs.~(\ref{eq:muzy}) and~(\ref{eq:zetap}) become inconsistent for larger values of~$H$ and~$\mu$, which persists for increased resolutions of the numerically calculated integrals.
We address this issue \textit{ad hoc} by numerically calculating a table, which maps the input parameters~$(H_\text{input},\mu_\mathrm{input})$ of Eq.~(\ref{eq:muzy}) to the parameters~$(H_\text{actual},\mu_\mathrm{actual})$ which best describe the resulting~$\zeta_p$ curves according to Eq.~(\ref{eq:zetap}).

The parameters used to mimic the Burgers data for each noise strength~$\sigma^2$ are summarized in Appendix~\ref{app:muzy}.

\begin{figure*}
    \centering
    \includegraphics[width=\linewidth]{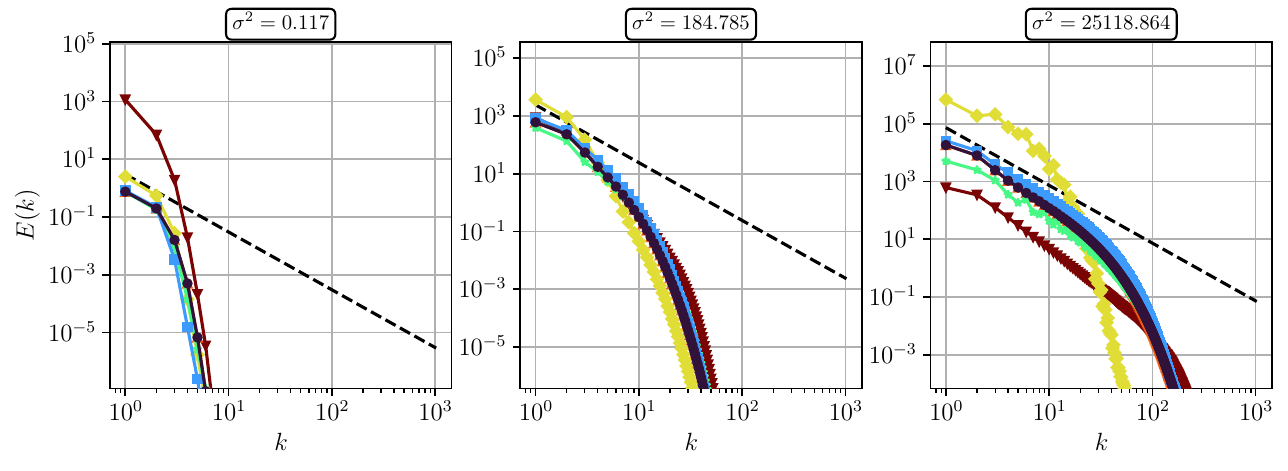}
    \includegraphics[width=\linewidth]{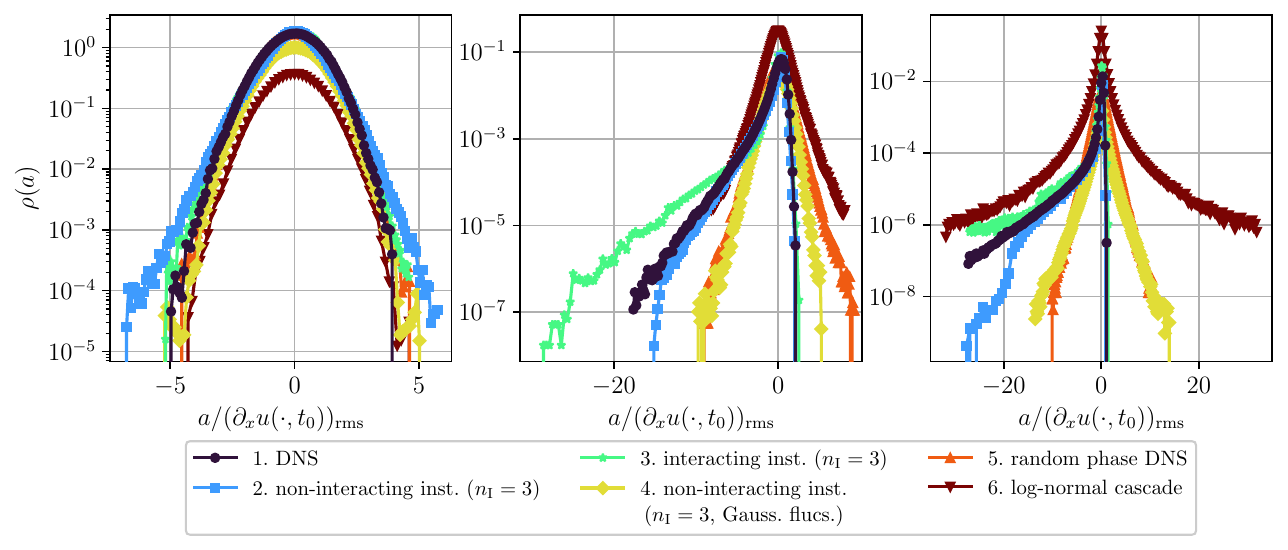}
    \caption{Energy spectrum $E(k) = \mathbb{E} \lvert\hat{u}_k \rvert^2 / 2$ (top row) and one-point gradient PDF $\rho(a)$ with $a = \partial_x u(0,0)$ (bottom row) at different noise strengths $\sigma^2$ or Reynolds numbers for the random field ensembles  listed in Sec.~\ref{sec:results}. The dashed black lines in the top row indicate $E(k) \propto k^{-2}$.}
    \label{fig:espec-grad-pdf}
\end{figure*}

\section{Results}
\label{sec:results}

In this section, we will compare Eulerian and Lagrangian statistical properties of the DNS fields to those of different synthetic ensembles as introduced in the previous sections. Concretely, we will consider the following six ensembles:
\begin{enumerate}
        \item DNS (ground truth)
        \item non-interacting canonical instantons with $n_{\text{I}}=3$
        \item interacting canonical instantons with $n_{\text{I}}=3$
        \item non-interacting canonical instantons with $n_{\text{I}}=3$ including Gaussian fluctuations around each instanton
        \item phase-randomized DNS fields
        \item log-normal cascade model
\end{enumerate}
For each of the ensembles listed above, we compute and analyze $N = 10^4$ realizations of the random fields at all 15 noise strengths $\sigma^2 \in [10^{-2}, 3 \cdot 10^5]$ from Sec.~\ref{sec:num-details}, or the corresponding equivalent ensembles. For a more detailed comparison of some statistical properties, we will use three representative noise strengths or (DNS) Reynolds numbers of $\sigma^2 \in \{1.17 \cdot 10^{-1}, 1.85 \cdot 10^2, 2.51 \cdot 10^4 \}$ and $\text{Re} \in \{1.03, 32.29, 192.21\}$, respectively.

\subsection{Eulerian statistics}
\label{sec:res-euler}
We start by discussing the snapshots of field realizations under the different ensembles in Fig.~\ref{fig:snapshots}. Note that for small $\sigma^2$ or $\text{Re}$ (top row), all random field ensembles necessarily yield similar results, as they effectively become Gaussian processes. Our main interest is in exploring how this changes as $\sigma^2$ increases.

The first column of Fig.~\ref{fig:snapshots} shows how field realizations from DNS of the Burgers Eq.~\eqref{eq:burgers} display steeper and steeper negative gradients as the noise variance or Reynolds number is increased, as expected. Since the noise covariance~$\chi$ used in the DNS is smooth and only excites a few Fourier modes, the velocity field of the Burgers DNS remains relatively smooth away from the quasi-shocks, even at high Reynolds numbers~\cite{bec-khanin:2007}. 

This picture is qualitatively reproduced by the instanton gas ensembles (columns 2 and 3 of Fig.~\ref{fig:snapshots}), as could be expected from Fig.~\ref{fig:inst-fields}. We see that, as remarked in Sec.~\ref{sec:grand-canonical}, the non-interacting ensemble always generates fields with (here) three instantons by construction. Depending on the realization of the independently and uniformly positions $x_{0j}$ of the instantons, this may give ``too many'' shocks at high Reynolds numbers, compared to the DNS fields. In contrast to this, the interacting ensemble, through volume exclusion effects of the interactions $\Delta S$ in Eq.~\eqref{eq:multi-inst-pdf}, can lead to only effectively one (Fig.~\ref{fig:snapshots}, $\sigma^2 = 25118.864$, red curve) or two (Fig.~\ref{fig:snapshots}, $\sigma^2 = 25118.864$, orange and blue curves) instantons being present, which appears in principle closer to DNS field realizations. The appearance of effective single-instanton configurations and regularly spaced two-instanton configurations in the interacting ensemble suggests that the effect of interactions is relatively strong in our particular setup. Going to larger domain sizes, or equivalently shorter forcing correlation lengths, would lead to a more dilute gas with a larger number of weakly interacting instantons as discussed in Sec.~\ref{sec:grand-canonical}.

Interestingly, adding Gaussian fluctuations around instantons incorporates additional slightly smaller-scale structures (column 4 of Fig.~\ref{fig:snapshots}), but does \textit{not} lead to a closer resemblance of the fields to DNS realizations at higher~$\sigma^2$, as is apparent from the large positive gradients at~$\sigma^2=25118.864$ that would never be observed in DNS of the Burgers Eq.~\eqref{eq:burgers}. This is perhaps surprising, given that we showed in Fig.~\ref{fig:ux-pdf-compare-inst} that including Gaussian fluctuations around the instanton in the theoretical estimation of the one-point gradient PDF~\eqref{eq:one-loop-grad} considerably improved the agreement with DNS. The prefactor calculation in Eq.~\eqref{eq:one-loop-grad} essentially incorporates random perturbations in the velocity field around the instanton configuration close to the shock location $x_0$, where $\partial_x \delta u(x_0, t_0) = 0$ vanishes for the fluctuations by construction~\cite{schorlepp-grafke-grauer:2021,schorlepp-tong-etal:2023}. It is intuitive that accounting for such slight shock deformations, compared to considering just the instanton, improves the prediction. The limit in which this picture is valid is when the gradient strength~$a$ of the instanton is rare for a given~$\sigma^2$, i.e.\ the effective smallness parameter is $h(a)/\sigma^2$ for some increasing function $h$ of $\lvert a \rvert$. In this limit, the Gaussian fluctuations~\eqref{eq:gauss-flucs} around the instanton are small and approximate the true, generally non-Gaussian conditional fluctuations~\cite{ebener-margazoglou-friedrich-etal:2019} with good accuracy.  In contrast to this, for high~$\sigma^2$ compared to~$\lvert a \rvert$ when the observable value is in the core of the distribution and not rare, what can happen (and indeed what we see in the bottom row of Fig.~\ref{fig:snapshots}) is that the fluctuations ``overpower'' the instanton. This means that we \textit{only} see the Gaussian random field given by the fluctuations, instead of the instanton, which can then be very far from the true conditional fluctuations~\cite{ebener-margazoglou-friedrich-etal:2019} at high~$\sigma^2$.

Because of this, we anticipate based on the snapshots in Fig.~\ref{fig:snapshots} that in the present example, adding independent Gaussian fluctuations around the individual instantons as described in Sec.~\ref{sec:gaussian-fluc} is actually detrimental to the random field sampling as~$\sigma^2$ increases. Sampling self-consistent Gaussian fluctuations around the \textit{sum} of instantons instead, introducing an effective noise strength in Eq.~\eqref{eq:gauss-flucs} that accounts for the smallness parameter~$h(a_j)/\sigma^2$, or making the Gaussian fluctuations part of the ansatz~\eqref{eq:u-sum-inst} inserted into the path integral, might all improve the results in the future, but we do not attempt this here. 

Lastly, the two ensembles without coherent structures (columns 5 and 6 of Fig.~\ref{fig:snapshots}) yield manifestly different realizations at high Re, displaying both strong positive and negative gradients at high $\sigma^2$, as well as much smaller-scale fluctuations and hence lower regularity than the DNS fields (while the energy spectrum, as a two-point quantity, of these fields is similar, or identical by construction, to the DNS fields).

The energy spectra $E(k) = \mathbb{E}[\lvert \hat{u}_k \rvert^2]/2$ for the different ensembles are shown in Fig.~\ref{fig:espec-grad-pdf} (top row). As $\sigma^2$ increases, more Fourier modes are excited in all ensembles, and the characteristic energy spectrum $E(k) \propto k^{-2}$ of Burgers turbulence begins to develop \cite{kida:1979,cheklov-yakhot:1995}. All synthetic field ensembles yield a good approximation to the energy spectrum of the DNS fields. Notably, already the non-interacting instanton gas gives a good approximation to the DNS spectrum, indicating once again that the approach captures the correct structures, i.e.\ the quasi-shocks, that are responsible for this scaling. Adding Gaussian fluctuations to these fields adds ``too much'' energy.

In the bottom row of Fig.~\ref{fig:espec-grad-pdf}, we show the one-point gradient PDFs $\rho(a)= \mathbb{E} \left[\delta(\partial_x u(x_0, t_0) - a) \right]$ of the different ensembles. The interacting and non-interacting instanton gas ensembles correctly reproduce the strongly asymmetric shape of the DNS PDF, with a much heavier, stretched exponential left tail. In fact, the non-interacting instantons give the best approximation here, while the interacting ensemble produces an even heavier left tail than the DNS fields. As commented above, this is probably due to the strong exclusion effects in our present setting, which push some instantons to larger negative gradients than in the non-interacting case. We also note that an important difference between the instanton prediction $\rho^{(0)}$ in Fig.~\ref{fig:ux-pdf-compare-inst}, and the PDFs shown here, is that the gradient at all points $x_0 \in [0, L_{\text{box}}]$ is considered here, whereas only the shock location itself is taken into account for the theoretical estimate in $\rho^{(0)}$ in Fig.~\ref{fig:ux-pdf-compare-inst}. That, and the inclusion of multiple instantons, leads to a much better agreement of the purely instantonic PDFs in Fig.~\ref{fig:espec-grad-pdf} with the DNS PDFs, compared to~$\rho^{(0)}$ itself. For the same reason as discussed above, including Gaussian fluctuations around the instantons destroys the asymmetry of the one-point gradient PDF in Fig.~\ref{fig:espec-grad-pdf} at high $\sigma^2$.

The structure-less synthetic fields 5.\ and 6.\ necessarily yield symmetric PDFs and hence fail to capture an essential property of the Burgers Eq.~\eqref{eq:burgers}, despite correctly reproducing the energy spectrum. The log-normal cascade model does at least produce heavy -- albeit symmetric -- tails at high $\sigma^2$. Although the phase-randomized data is expected to be Gaussian, we observe exponential tails of the gradient distributions. This is due to high variability of variances of the DNS fields, so while individual phase-randomized fields have Gaussian distributions, the ensemble average shown here constitutes a mixture of Gaussians.

\begin{figure*}
    \centering
    \includegraphics[width=\linewidth]{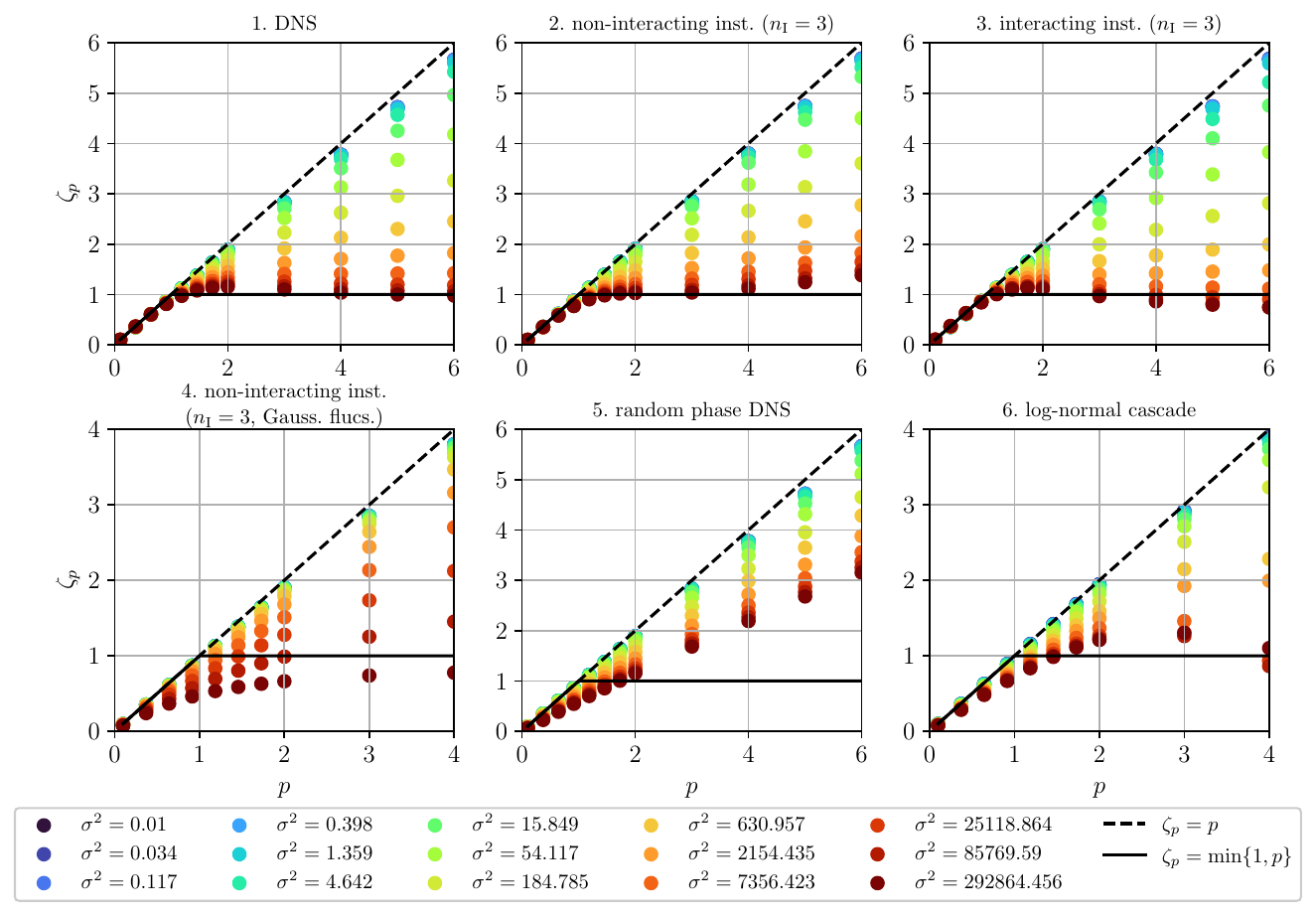}
    \caption{Structure function exponents $\zeta_p$, determined from log-log linear fits to $S_p(r) \propto r^{\zeta_p}$, as defined in Eq.~\eqref{eq:struc-func}, in the inertial range, for the random field ensembles at different noise strengths $\sigma^2$ or Reynolds numbers. The linear fits were performed in the range $r \in [0.1, 1]$ for all ensembles and noise variances $\sigma^2$. In ensembles 4.\ and 6., we show only exponents until $p = 4$, as higher-order structure functions did not display any clear scaling behavior in the fit range.}
    \label{fig:struc-comp}
\end{figure*}

Lastly, we turn to higher-order two-point quantities, in the form of structure functions
\begin{align}
    S_p(r) := \mathbb{E} \left[\lvert u(x+r) - u(x) \rvert^p\right]
    \label{eq:struc-func}
\end{align}
for $p > 0$. At high $\sigma^2$, the Burgers Eq.~\eqref{eq:burgers} is known to display the well-known shock scaling $S_p(r) \propto r^{\zeta_p}$ with $\zeta_p = \min \{1,p\}$, for $r$ in the inertial range \cite{bec-khanin:2007}. For small~$\sigma^2$ and~$r$, and a smooth Gaussian random field, it is easy to see by Taylor expansion that one has $S_p(r) \propto r^{\zeta_p}$ with $\zeta_p = p$ instead. We fix a fit range $r \in [0.1, 1]$, and perform a linear regression of $\log S_p$ as a function of $\log r$ in that range for the different random field ensembles. We remark that there are more sophisticated methods to estimate the inertial range and structure function exponents~\cite{benzi-ciliberto-etal:1993,grauer-homann-pinton:2012}, but we choose the simplest possible approach here for a fair comparison of the different ensembles~\footnote{In reality, the inertial range, if present at all, is for example not always of the same size as the Reynolds number varies}.

Even with this simple approach, we see that the Burgers DNS fields interpolate between $\zeta_p = p$ and  $\zeta_p = \min \{1,p\}$ as theoretically expected. 
Note that, since the fields are quite smooth, estimation of high moments up to order $p=6$, with clean scaling, are possible here. Again, the behavior of the structure function exponents is reproduced quite well by the instanton gas ensembles~2.\ and~3., while the instanton gas with Gaussian fluctuations does not converge to $\zeta_p = \min \{1,p\}$ at high $\sigma^2$.
The random phase DNS fields also show linear scaling $\zeta_p \propto p$ for small $\sigma^2$, converging to $\zeta_p = p/2 = (-\alpha-1)/2$ as $\sigma^2$ increases, which is the expected scaling for a monofractal field with spectral power-law slope $\alpha=-2$ \cite{frisch:1995}.
The log-normal cascade model shows the characteristic multifractal quadratic scaling described in Section~\ref{sec:lognormal-cascade}.

\subsection{Particle statistics}
\label{sec:res-lagrange}

In this subsection, we analyze Lagrangian statistics of the field ensembles from the previous subsection in two different setups. For this purpose, we consider the fields $u = u(x)$ as frozen while particles move through them, and also normalize all fields to unit root-mean-square velocity of their respective ensemble.

\subsubsection{In one dimension}

\begin{figure}
    \centering
    \includegraphics[width=\linewidth]{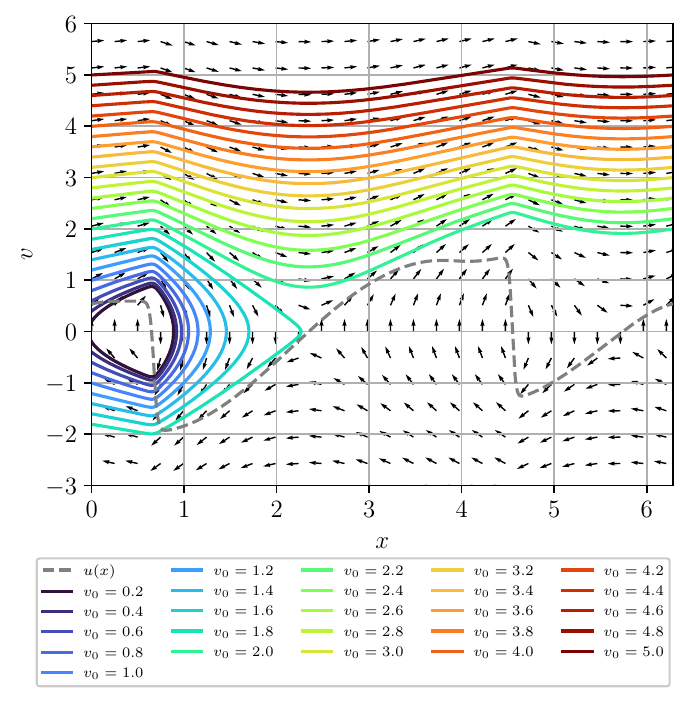}
    \caption{Particle propagation in $(x,v)$-phase space according to Eq.~\eqref{eq:particles} for different initial velocities $v_0$, illustrating the possibility of particles being captured by shocks. The field $u(x)$ in this figure (dashed gray line) is sampled from the non-interacting canonical instanton ensemble with $n_{\text{I}} = 3$ at $\sigma^2 = 25118.864$. Note that since the aspect ratio of the plot is not 1, the arrows indicating the right-hand side of Eq.~\eqref{eq:particles} do not appear exactly tangent to the trajectories.}  
    \label{fig:prop-time-example}
\end{figure}

First, as the simplest possibility motivated by future applications to cosmic ray propagation, in our one-dimensional example of the Burgers Eq.~\eqref{eq:burgers}, we consider unit mass particles with position $x(t)$ and velocity $v(t)$, on which the field $u$ acts as a force:
    \begin{align}
\begin{cases}
    \frac{\mathrm{d} x}{ \mathrm{d} t} = v(t)\,, \quad & x(0) = 0\,,\\
    \frac{\mathrm{d} v}{ \mathrm{d} t} = u(x(t))\,, \quad & v(0) = v_0 \,,
\end{cases}
        \label{eq:particles}
    \end{align}
Here, we start the particles at the left end $x = 0$ of the box $[0, L_{\text{box}}]$, and vary their initial velocity $v_0 \in [0.2,5]$. Note that this one-dimensional situation is necessarily quite different from the magnetic Lorentz force in three dimensions, in that the force in Eq.~\eqref{eq:particles} can change the magnitude of the particles' velocities, and is hence more comparable to an electric field. Nevertheless, particles can still be ``captured'' (similar to magnetic mirrors in MHD) or accelerated by shock-like structures. The setup is illustrated as a phase portrait in Fig.~\ref{fig:prop-time-example}. We see that the quasi-shocks act as centers in the phase portrait. If the particle has a small initial velocity $v_0$ at $x=0$, it is captured by the first shock. If instead it starts above the separatrix, it will traverse the domain, but the time it takes will still be affected by the structure of the random field (less so as $v_0$ increases).

For each initial velocity $v_0 > 0$, we determine whether the particle can cross the box, and if yes, within which time $t_{\text{arrival}}$ compared to the ballistic propagation time for $u = 0$, which is $t_{\text{ballistic}} = L_{\text{box}} / v_0$. For this purpose, we integrate Eq.~\eqref{eq:particles} numerically with a time-adaptive solver, interpolating $u$ linearly between grid points, and until a maximum time of $2.5 \cdot t_{\text{ballistic}}$.

\begin{figure*}
    \centering
\includegraphics[width=\linewidth]{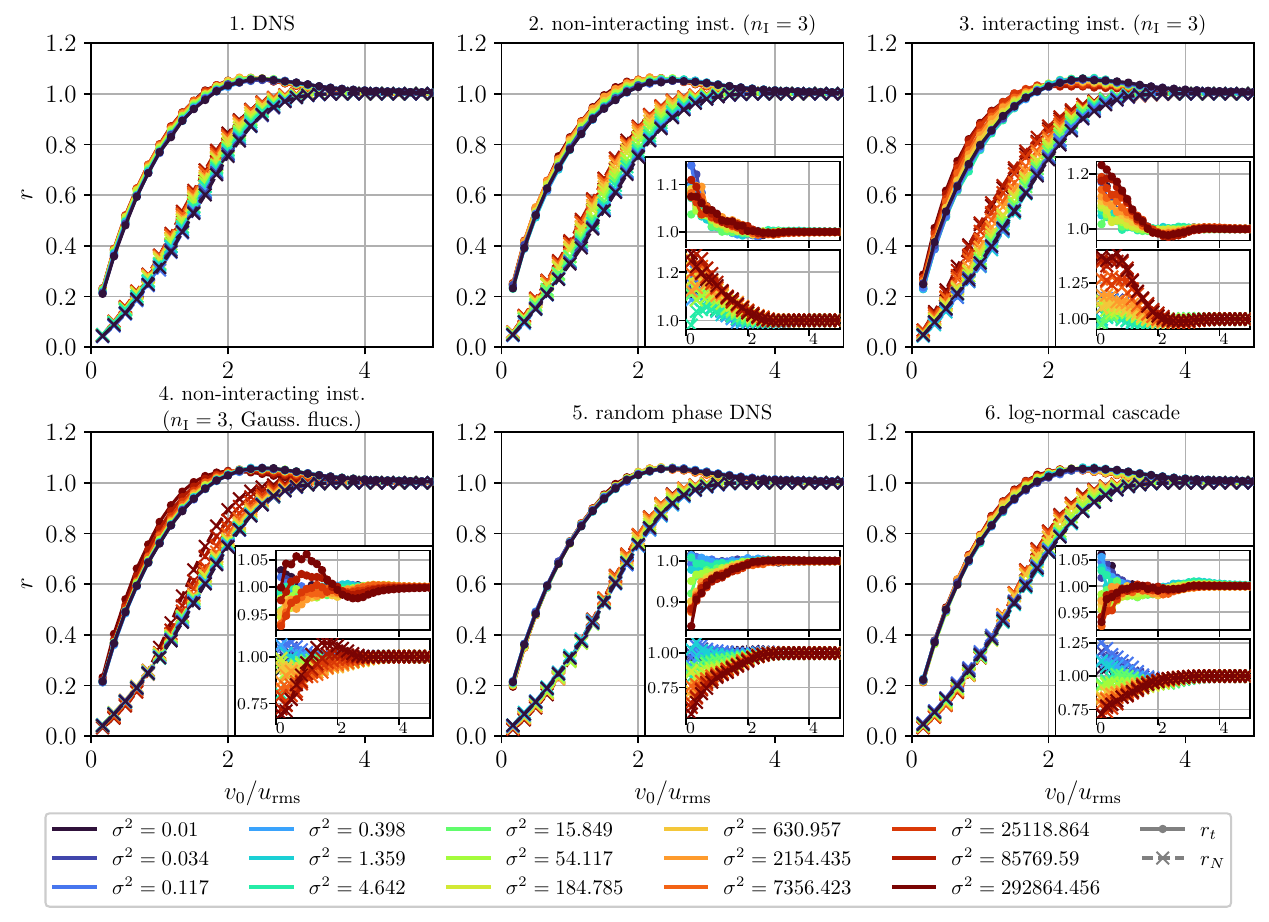}
    \caption{Comparison of normalized average time for particles to cross the domain $r_t = \mathbb{E}\left[t_{\text{arrival}} \mid t_{\text{arrival}} < \infty \right] /  t_{\text{ballistic}}$ (dots), where $t_{\text{ballistic}} = L_{\text{box}} / v_0$, and fraction of particles that manage to cross the domain $r_N = N_{\text{arrival}} / N$ (crosses) from $x=0$ to $x=L_{\text{box}}$, for the different random field ensembles when varying the noise strength $\sigma^2$ or Reynolds number. All $N = 10^4$ fields per ensemble are normalized to ensemble unit root-mean square velocity for this experiment, and $r_t, r_N$ are shown as a function of the initial velocity $v_0$, cf.\ Fig.~\ref{fig:prop-time-example}. The insets show the quotient of the respective ensemble's $r_t$ and $r_N$, relative to the DNS result.}
    \label{fig:prop-times}
\end{figure*}

We consider all field ensembles from the previous subsection with $N = 10^4$ field realizations each, and for each~$v_0$ compute the fraction of particles that do cross the domain $r_N = N_{\text{arrival}} / N$, as well as the average time needed for those that arrive, compared to the ballistic time: $r_t = \mathbb{E}\left[t_{\text{arrival}} \mid t_{\text{arrival}} < \infty \right] /  t_{\text{ballistic}}$. The results are summarized in Fig.~\ref{fig:prop-times}. As expected, for all random field ensembles, in the limit of large initial velocity $v_0 \gg 1$, when almost all particles pass through the domain and are mostly unaffected by the force field, we see that $r_t, r_N \to 1$, and similarly, for the opposite case of slow particles $0 < v_0 \ll 1$, we have $r_t, r_N \to 0$. Furthermore, since all fields are normalized to root-mean-square velocity~$1$ for this experiment, and all ensembles 1.\ to 6.\ listed above become smooth and Gaussian for small~$\sigma^2$, the general shape of the~$r_t$ and~$r_N$ curves is similar among all ensembles, and differences will mainly be visible in the Reynolds number dependent change of the curves as~$\sigma^2$ increases.

The fraction of particles that pass through the domain~$r_N$ is necessarily a monotonically increasing function of~$v_0$. In contrast to this, the (normalized) average time to pass through the domain~$r_t$ can be seen to be a non-monotonic function of~$v_0$, and particles below $v_0 \approx 1.8$ are on average slowed down by the force field~$u$, whereas particles with larger~$v_0$ are on average faster than without the force field.  We do see a qualitative difference in the Reynolds number dependence of the curves between the DNS fields and the purely instantonic ensembles~2.\ and~3.\ on the one hand, and the remaining ensembles on the other. For fixed~$v_0$ and as~$\sigma^2$ increases, the fraction~$r_N$ increases monotonically in the DNS fields, and this is correctly reproduced by the instanton gas ensembles without fluctuations. A reason for this could be that quasi-shocks become more localized as~$\sigma^2$ increases, and hence particles being trapped is less likely. In contrast to this, the structureless ensembles 5.\ and 6., as well as the instanton gas with Gaussian fluctuations show a different behavior, where below a critical velocity~$v_{0,\text{c}}$, $r_N$ actually decreases as a function of~$\sigma^2$, and increases above~$v_{0,\text{c}}$. Again, the reason for the very different results for the instanton gas with Gaussian fluctuations, compared to those of the instanton gas without fluctuations, is that due to the fluctuations being ``too strong'' away from the instanton locations~$x_{0j}$, all that the particles see is the Gaussian random field given by the fluctuations. All in all, we take these results as a first indication that the instanton gas approximation succeeds in capturing essential properties of the DNS fields for particle propagation, that are missing in other synthetic fields.

\subsubsection{Cosmic ray statistics}
We now move on to a second, more realistic setup closer to cosmic ray propagation in MHD fields.
For this purpose, we construct three-dimensional, divergence-free magnetic fields as slab turbulence \cite{mertsch:2020,els:2024}
\begin{equation}
    \bm{B}(x,y,z)=\left(1, 0, s\,(\partial_xu)(x)\right)^\top,
    \label{eq:magnetic-field}
\end{equation}
where we take the gradient of our one-dimensional fields~$\partial_xu$ as turbulent fluctuations in $z$-direction with relative strength~$s$ along a uniform background field $B_0=1$ in $x$-direction.
This simple toy model, which mimics the anisotropic nature of MHD turbulence, provides us with an easy way to transfer our various turbulence models to the manifestly three-dimensional motion of charged particles subjected to the Lorentz force.
The corresponding equations of motion are given by
\begin{equation}
    \begin{cases}
        \frac{\mathrm{d}\bm{r}}{\mathrm{d}t}=\bm{v}(t), & \bm{r}(0)=(0,0,0)^\top, \\
        \frac{\mathrm{d}\bm{v}}{\mathrm{d}t}=\frac{q}{m}\bm{v}(t)\times\bm{B}\!\left(\bm{r}(t)\right), & \bm{v}(0)=\bm{v}_0,
    \end{cases}
\end{equation}
where test particles are parametrized by their initial velocity $\bm{v}_0$ and their charge-to-mass ratio $q/m$.
The particles perform a gyro motion about the $x$-direction with a maximal gyro radius $r_{\text{g,max}}=v_0m/q$.
Since we neglect the influence of an electric field, test particle energies are conserved.
We also record the pitch-angle cosine
\begin{align}
    \mu=\hat{\bm{v}}\cdot\hat{\bm{B}}
\end{align}
(not to be confused with the intermittency parameter~$\mu$ of the log-normal cascade model from Sec.~\ref{sec:lognormal-cascade}) as a useful indicator for particle motion parallel to the uniform background field.
To integrate this equation numerically, we employ a volume-preserving Boris integrator \cite{boris:1971,ripperda:2018}, with an adaptive step size based on a simple predictor-corrector scheme. The magnetic field is interpolated at the particle location by evaluating a local third-order Taylor expansion.
\begin{figure}
    \centering
    \includegraphics[width=\linewidth]{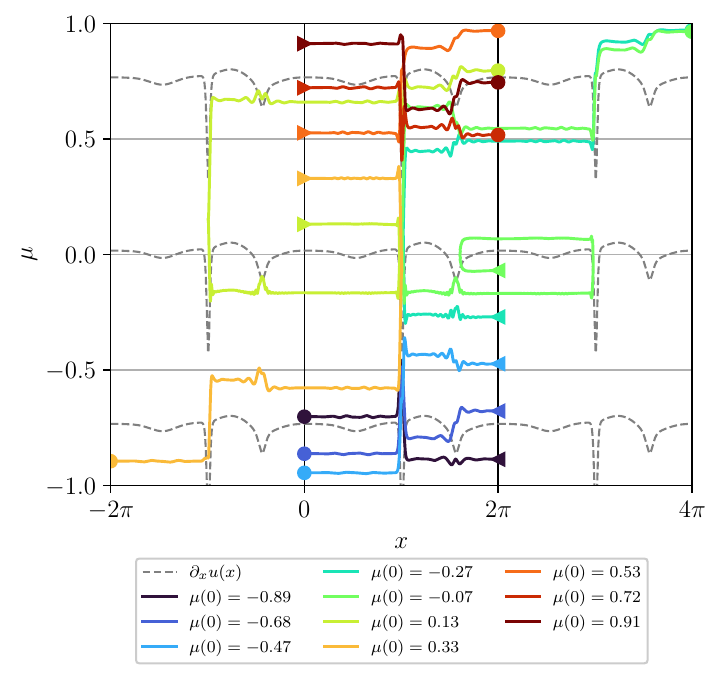}
    \caption{Particle trajectories~$(x(t),\mu(t))$ with various initial pitch-angle cosine values~$\mu(0)$ in an example field realization~$\bm{B}(x)=(1,0,0.1\,\partial_xu(x)/(\partial_xu)_\mathrm{rms})^\top$, where~$u(x)$ is sampled from the Burgers DNS ensemble with~$\sigma^2=25118.864$. The sign of~$\mu(t)$ indicates the current travel direction of a particle ($\mu>0$: rightwards, $\mu<0$: leftwards). Initial positions are marked with~$\triangleright$ for~$\mu(0)>0$ and~$\triangleleft$ for~$\mu(0)<0$, and final positions are marked with~$\circ$.}
    \label{fig:3d-prop-sketch}
\end{figure}
{Sample trajectories in one realization of a magnetic field described by Eq.~\eqref{eq:magnetic-field} are shown in Fig.~\ref{fig:3d-prop-sketch}. Here, periodic boundary conditions are imposed to account for reversals of the travel direction. The turbulent component of the magnetic field given by~$\partial_xu(x)$ is highly structured, consisting of long and weak segments, and sudden and intense spikes. Particles follow along the weak segments mostly undisturbed, but exhibit sudden jumps in~$\mu$ when encountering intense spikes. These jumps may reverse the sign of~$\mu$ and constellations of spikes may confine particles between them. We note that these reversals of direction are not classical mirror events, because the magnetic moment is not conserved during these events, which are characterized by short time scales.}

To avoid misleading statistics, particles should not travel repeatedly through the same periodic domain with length $2\pi$, so we restrict particle trajectories to $|x_t-x_0|<2\pi$.
Consequently, long-time diffusive behavior cannot be studied, however the short-time evolution already allows interesting insights into the basic scattering mechanisms of the various turbulence models.

\begin{figure*}
     \centering
\includegraphics[width=\linewidth]{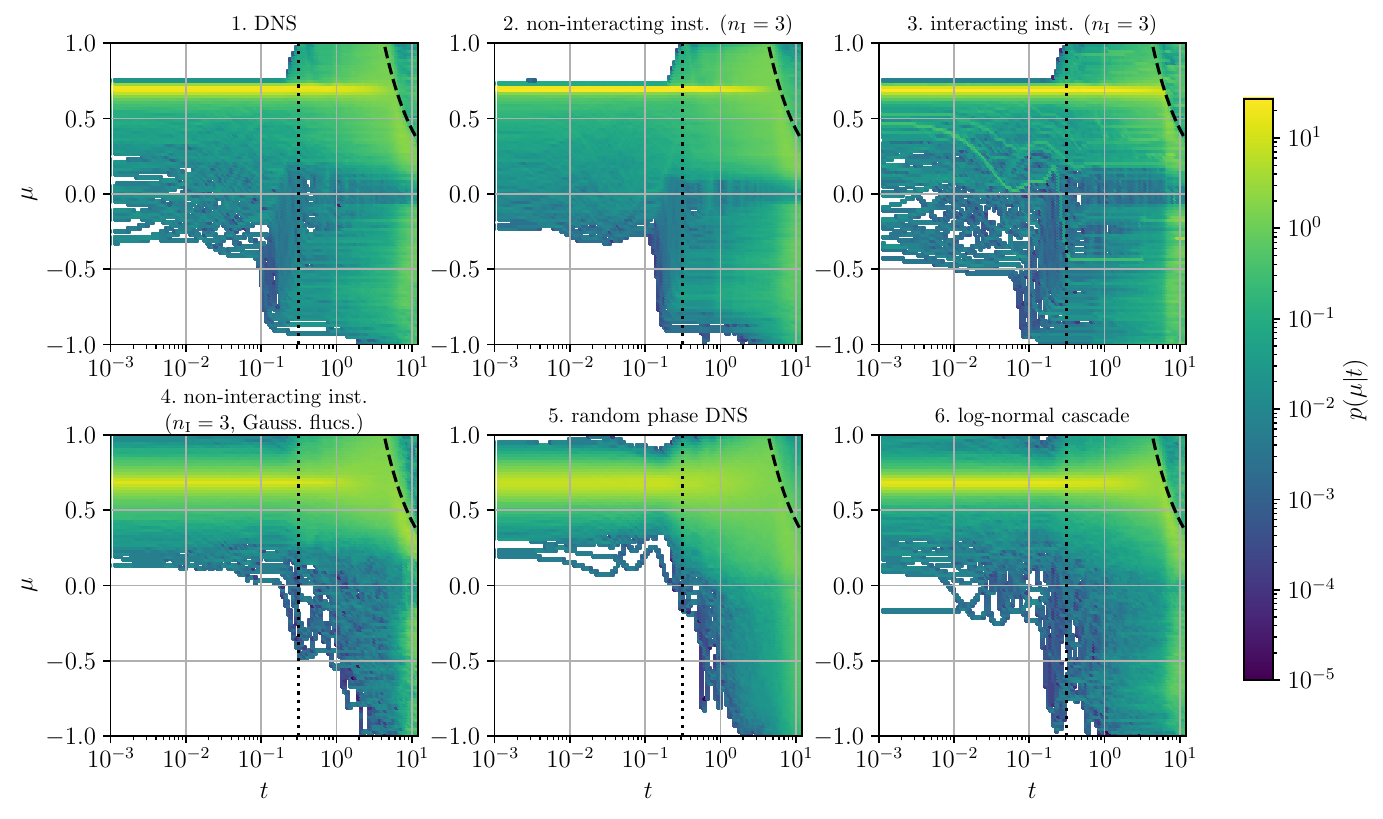}
    \caption{Time-dependent distributions of the pitch-angle cosine~$p(\mu|t)$ for our turbulence models. Particles are initialized with~$\mu_0=0.684$ or~$\arccos\mu_0\approx46.85^\circ$ with respect to the background magnetic field. 
    The dotted line indicates a half gyro period~$T_g/2=\pi m/q$, and the dashed line indicates the expected box-crossing time~$t_\text{box}=2\pi\mu_0/\mu$. 
    Note the similarity of the distributions shown in the top row, indicating the efficiency of the synthetic turbulence models constructed from the interacting and even non-interacting instanton gas.}
    \label{fig:mu-pdfs}
\end{figure*}

We illustratively study the effect of our various turbulence models on the particle dynamics on the basis of the temporal evolution of the pitch-angle cosine distribution~$p(\mu|t,\mu_0)$ for a single parameter set.
We set the strength parameter to $s=0.1/\left(\partial_xu\right)_\mathrm{rms}$, corresponding to a turbulence strength of $\delta B/B=0.1$, and we set the charge-to-mass ratio to the moderate value $q/m=10$, where the resulting dynamics are equally shaped by particle momentum and turbulent fluctuations. 
Particle velocities are initialized with $v=1$ and a pitch-angle cosine $\tilde{\mu}_0=\hat{\bm{v}}\cdot\hat{\bm{B}}_0$ with respect to the uniform background field (the actual pitch-angle cosine distribution at $t=0$ then shows some natural spread due to the turbulent fluctuations). 
The results are displayed in Fig.~\ref{fig:mu-pdfs}.

The case of Burgers DNS exhibits an intense beam at~$\mu_0$, which remains recognizable until the expected box-crossing time.
This beam is formed by particles which travel along extended quiet regions of the Burgers turbulence and only scatter intermittently upon encountering strong shocks.
These scattering events exhibit high amplitudes (including sign reversals) and are non-adiabatic as shown by the sudden broadening of the distribution at~$t\sim T_g/2$, i.e.~the expected time scale where turbulent fluctuations dominate over the particle momentum.
The results also show an initial spread of the distribution towards smaller~$\mu$, which is due to particles with initial positions in close vicinity to intense shocks.
The non-interacting and interacting instanton models resemble this transport behavior quite remarkably.

On the other hand, the remaining turbulence models (consisting of non-interacting instanton gas with Gaussian fluctuations, random phase data and the log-normal cascade model) are qualitatively similar to each other, but show distinct differences to the DNS and pure instanton cases.
Specifically, the initial distribution shows symmetric spread around~$\mu_0$ due to small space-filling fluctuations present in these models, leading to a less clearly defined beam.
Also the scattering events experienced by particles at~$t\gtrsim T_g/2$ are more frequent and less intense, leading to a diffusion-like behavior.
Since ensembles 4 and 6 still exhibit intermittent fluctuations, the initial beam is slightly more pronounced compared to ensemble 5 (random phases).
Note that the random-phase case corresponds to the well-studied quasi linear theory of cosmic ray transport, which prohibits scattering through~$\mu=0$, i.e.\ reversals or mirroring, in the magneto-static case \cite{mertsch:2020}.
The observed emergent distribution at~$\mu<0$ at later times is due to numerical inaccuracies.

\section{Conclusion}
\label{sec:concl}

In this paper, we have presented a coherent-structure based approach to synthesizing turbulent random fields. In contrast to many previous works in this direction, by starting from the field-theoretic formulation of turbulence and approximating the path integral with a suitable ansatz of instanton solutions, this yields a principled approach that uses both coherent structures and target distributions of their parameters that can be derived from the underlying equations of motion of the fluid. For concreteness, we have illustrated the method in one-dimensional forced Burgers turbulence. We observe that a superposition of instantons reproduces the Eulerian and Lagrangian statistics of DNS fields surprisingly well. Intuitively, this is due to the fact that the DNS fields are composed of smooth ramps and (quasi)shocks at high Reynolds numbers, and the instanton configurations precisely yield those structures. Importantly, however, both the instanton configurations to superimpose (which are special solutions of the forced Burgers equation) and their random distribution are not arbitrarily chosen, such that the method can be generalized to other settings.

Throughout all numerical experiments, we have seen that a direct superposition of instantons, with or without interactions $\Delta S$ among them, gave a more accurate representation of DNS statistics than Gaussian random phase fields or a lognormal cascade model without coherent structures. Specifically, while all ensembles display comparable energy spectra,  deviations between the structureless random fields and DNS fields become apparent when studying structure function exponents, one-point gradient PDFs, or particle propagation statistics in one-dimensional force fields (for particle acceleration and confinement) or three-dimensional slab-like magnetic fields (for the pitch angle distribution and reversal events).

Among the different instanton ensembles considered here, already a canonical ensemble of instantons without interactions or fluctuations seems be successful in capturing the DNS statistics quite well. This will not necessarily be the case in other systems beyond the one-dimensional Burgers equation. We have seen that including interactions among instantons, which can incorporate volume exclusion effects and effectively remove some of the instantons, only yields small modifications of the Eulerian and Lagrangian statistics we analyzed, and in some cases even results in worse agreement with DNS than in the non-interacting case. Similarly, we have discussed the inclusion of independent Gaussian fluctuations around individual instantons. While Gaussian fluctuations considerably improve the theoretical one-point gradient PDF prediction as shown in Figs.~\ref{fig:ux-pdf-compare-inst} and~\ref{fig:theta} even at high noise strengths~$\sigma^2$, they tend to overpower the instantons in the random fields for instanton observable values~$a_j$ that are in the core of the gradient PDF, such that the resulting fields look almost Gaussian and hence less and less like DNS fields at high~$\sigma^2$.

We have pointed out possible modifications and improvements of the presented methods along the way, which could help to ameliorate this and other points. We mention here the possibility of using approximate interactions between instantons for faster sampling, using a truly grand canonical ensemble of instantons, adding self-consistent noise corrections to interactions between instantons as proposed in Ref.~\cite{millo-faccioli-scorzato:2010}, and further sampling consistent Gaussian or even higher-order fluctuations around the sum of instantons. The field-theoretic approach to turbulence that we followed here allows for a systematic way of exploring these ideas.

\section{Outlook: 2D MHD turbulence}
\label{sec:outl}

As an outlook on higher-dimensional problems, in order to synthesize turbulent fields via an instanton gas approximation in real turbulence, in near future work we envisage the case of two-dimensional incompressible MHD turbulence~\cite{biskamp:2003,schekochihin:2022}. This serves as a bridge to synthesize three-dimensional MHD turbulence on very large astrophysical domains, to address the issue of cosmic ray transport in highly intermittent turbulence containing a hierarchy of coherent structures.

\begin{figure*}
    \centering
    \includegraphics[width=\linewidth]{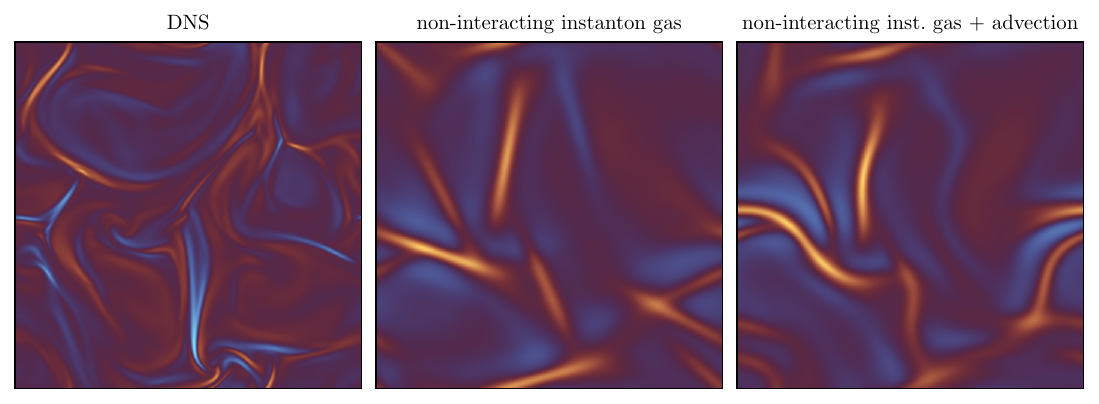}
    \caption{Snapshots of the current density field $j(x,y) = (\nabla \times \bm{B})_z(x,y)$ for the stochastically forced two-dimensional MHD equations. Left: snapshot from DNS. Center: non-interacting instanton gas configuration. Right: non-interacting instanton gas plus additional passive advection by large-scale random field.}
    \label{fig:MHDinstantonGas}
\end{figure*}

We show a preliminary proof of concept in Fig.~\ref{fig:MHDinstantonGas}. Here, instantons were calculated for the two-dimensional incompressible MHD equations with the current density $j = (\nabla \times \bm{B})_z = \partial_x B_y - \partial_y B_x$ at one point in the domain as observable, using an optimization procedure similar to that described in this paper and following Ref.~\cite{schorlepp-grafke-may-etal:2022}. As a placeholder for the sampling procedures described in this paper, we used the one-point current density PDF obtained from a pseudo-spectral DNS run of the stochastically forced 2D MHD equations (see Fig.~\ref{fig:MHDinstantonGas} (left) for a DNS field snapshot) to independently draw the different strengths of the current density of the instantons~\footnote{Note that this is \textit{not} consistent with Eq.~\eqref{eq:pdf-noninter} for non-interacting instantons, which instead includes the instanton action and Gramian for sampling the observable values. For the purpose of this qualitative outlook, we content ourselves with drawing from the one-point DNS PDF without Gramian, since it requires computing fewer instantons to generate a single snapshot -- we draw observable values from the DNS PDF, and compute instantons just for those values.}.
At this point, we would like to note that in general, spontaneous breaking of continuous symmetries of the instanton solutions can occur, cf.~\cite{schorlepp-grafke-may-etal:2022}. This is also the case for the present example of 2D MHD turbulence, where rotational symmetry is broken by the instanton solutions for the current density. The parameters that label the submanifold of possible instanton solutions (here, a single angle) then have to be included as additional collective coordinates for the instanton gas.

Therefore, each instanton is randomly rotated and shifted, before being summed to form the final turbulent current density field in Fig.~\ref{fig:MHDinstantonGas} (center). In addition, Fig.~\ref{fig:MHDinstantonGas} (right) shows a turbulent field resulting from subsequent passive advection of the instanton gas, motivated by the findings discussed in Ref.~\cite{laval-dubrulle-nazarenko:1999}. To achieve this, a large-scale Gaussian random velocity field with a few Fourier modes ($\lVert k \rVert < 6$) was used to advect the stream function corresponding to the instanton gas realization from Fig.~\ref{fig:MHDinstantonGas} (center). Many aspects still require further investigation, e.g.\ whether there is a need for self-avoiding superposition of current sheets, whether there are further constraints besides numerical issues to produce fast methods applicable on large-scale domains, and whether the effect of passive advection and entanglement of structures can be achieved in a more self-consistent way, e.g.\ through interactions and fluctuations. Nevertheless, on a qualitative level, the field realization in Fig.~\ref{fig:MHDinstantonGas} (right) already displays similar structures as the DNS snapshot.

\section*{Acknowledgments}
K.K., J.L., and R.G.\ acknowledge support from the German Research Foundation DFG within the Collaborative Research Center SFB1491. T.S.\ acknowledges support from the NSF grant DMS-2012548 and the PSC-CUNY grant TRADB-55-275.

\bigskip

\appendix
\section{Estimation of hyperparameters for the log-normal continuous cascade model}\label{app:muzy}
Table~\ref{tab:params} summarizes the parameters employed by the log-normal cascade model Eq.~(\ref{eq:muzy}) to mimic the Burgers DNS data at low orders~$p<5$.

\begin{table*}
    \centering
    \caption{Employed parameters of the log-normal continuous cascade model for each DNS noise strength $\sigma^2$. The parameters $H_{\text{input}}$ and $\mu_{\text{input}}$ are used as input parameters to Eq.~(\ref{eq:muzy}) for the purpose of generating the synthetic random fields. The resulting $\zeta_p$ curves are best described by Eq.~(\ref{eq:zetap}) with the parameters $H_{\text{actual}}$ and $\mu_{\text{actual}}$. The small-scale cutoff $\ell$ in Eq.~\eqref{eq:muzy} controls the spectral cutoff behavior.}
    \label{tab:params}
    \begin{ruledtabular}
    \bgroup
    \def\arraystretch{1.2}
    \begin{tabular}{cccccc}
         $\sigma^2$ & $\ell$ & $H_\mathrm{actual}$ & $H_\mathrm{input}$ & $\mu_\mathrm{actual}$ & $\mu_\mathrm{input}$\\[2pt]
         \hline\\[-10pt]
 0.010 & 0.539 & 0.978 & 1.99998 & $1.32\times10^{-5}$ & 0.000 \\
 0.034 & 0.511 & 0.978 & 1.99994 & $4.16\times10^{-5}$ & 0.001 \\
 0.117 & 0.480 & 0.978 & 1.99985 & $6.18\times10^{-5}$ & 0.002 \\
 0.398 & 0.415 & 0.978 & 1.99959 & $0.000352$ & 0.005 \\
 1.359 & 0.327 & 0.978 & 1.99888 & $0.00135$ & 0.012 \\
 4.642 & 0.218 & 0.978 & 1.99696 & $0.00512$ & 0.034 \\
15.849 & 0.141 & 0.977 & 1.99179 & $0.0195$ & 0.091 \\
54.117 & 0.098 & 0.977 & 1.97817 & $0.0599$ & 0.241 \\
184.785 & 0.063 & 0.976 & 1.94392 & $0.138$ & 0.559 \\
630.957 & 0.036 & 0.970 & 1.86736 & $0.24$ & 0.876 \\
2154.435 & 0.023 & 0.957 & 1.73351 & $0.344$ & 0.977 \\
7356.423 & 0.016 & 0.939 & 1.57619 & $0.425$ & 0.994 \\
25118.864 & 0.011 & 0.920 & 1.45862 & $0.477$ & 0.996 \\
85769.590 & 0.008 & 0.904 & 1.39708 & $0.505$ & 0.997 \\
292864.456 & 0.006 & 0.891 & 1.37077 & $0.52$ & 0.997 \\
    \end{tabular}
    \egroup
    \end{ruledtabular}
\end{table*}

\bibliography{bib.bib}

\end{document}